\documentclass[aps,prx,amsmath,superscriptaddress,twocolumn,bibnotes,longbibliography,floatfix]{revtex4-2}
\usepackage{amsmath}
\usepackage{amsfonts}
\usepackage{graphicx}
\usepackage{color}
\usepackage{dsfont}
\usepackage{verbatim}
\usepackage{mathtools}
\usepackage[normalem]{ulem}
\usepackage{comment}
\usepackage{stackrel}
\usepackage{bm}
\usepackage{booktabs}
\usepackage{comment}

\usepackage[colorlinks=true, 
linkcolor=blue, 
urlcolor=blue,
citecolor=blue]{hyperref}
\usepackage{orcidlink}
\usepackage[linesnumbered]{algorithm2e}

\RestyleAlgo{ruled}
\SetKwComment{Comment}{/* }{ */}
\definecolor{myblue}{rgb}{.93, .93, 1}

\newcommand{\bsub}{\begin{subequations}}
	\newcommand{\esub}{\end{subequations}}

\begin{document}
	
	\title{Composite helical edges from Abelian fractional topological insulators}
	
	\author{Yang-Zhi~Chou~\orcidlink{0000-0001-7955-0918}}\email{yzchou@umd.edu}
	\affiliation{Condensed Matter Theory Center and Joint Quantum Institute, Department of Physics, University of Maryland, College Park, Maryland 20742, USA}

	\author{Sankar Das~Sarma~\orcidlink{0000-0002-0439-986X}}
	\affiliation{Condensed Matter Theory Center and Joint Quantum Institute, Department of Physics, University of Maryland, College Park, Maryland 20742, USA}
	
	\date{\today}

	\begin{abstract}
		We study low-temperature phases and edge-state conductance of an interacting composite $(1+1/n)$ Abelian helical edge state made of a regular helical liquid carrying charge $e$ and a (fractionalized) helical liquid carrying charge $e/n$. A systematic framework is developed for these composite $(1+1/n)$ Abelian helical edge states with $n=1,2,3$, incorporating symmetry-allowed interaction and disorder. For $n=1$, the system is made of two regular helical Luttinger liquids. For $n=2$, the composite edge state consists of a regular helical Luttinger liquid and a fractional topological insulator (the Abelian $Z_4$ topological order) edge state arising from half-filled conjugated Chern bands. The composite edge state with $n=2$ is pertinent to the recent twisted MoTe$_2$ experiment at $\nu_{\text{tot}}=3$, suggesting a possible fractional topological insulator with an edge-state conductance $\frac{3}{2}\frac{e^2}{h}$. For $n=3$, we consider a regular helical Luttinger liquid and a pair of time-reversal Laughlin $\nu=1/3$ fractional quantum Hall edge states. Using bosonization, we identify relevant perturbations and construct generic phase diagrams in the presence of \textit{weak} Rashba spin-orbit coupling. In addition to a phase of free bosons, we find a time-reversal symmetry-breaking localized insulator, two perfect positive drag phases (due to different mechanisms), a perfect negative drag phase (for $n=2,3$), a time-reversal symmetric Anderson localization (only for $n=1$), and a disorder-dominated metallic phase analogous to the $\nu=2/3$ disordered fractional quantum Hall edges (only for $n=3$). We further compute the two-terminal edge-state conductance, the primary experimental characterization for the (fractional) topological insulator. Remarkably, the negative drag phase gives rise to an unusual edge-state conductance, $(1-1/n)\frac{e^2}{h}$, which is not directly associated with the filling factor. We further investigate the effect of an applied in-plane magnetic field. For $n>1$, the applied magnetic field can result in a phase with localization only in the regular helical liquid channel, and the edge-state conductance is $\frac{1}{n}\frac{e^2}{h}$, providing another testable signature. Our work establishes a systematic understanding of the composite $(1+1/n)$ Abelian helical edge, paving the way for future experimental and theoretical studies.
	\end{abstract}
	
	\maketitle
	
	\section{Introduction}
	
	A topological insulator (TI) is a gapped system exhibiting nontrivial symmetry-protected edge/surface states that avoid Anderson localization \cite{HasanMZ2010a,QiXL2011,HasanMZ2011,SenthilT2015,BansilA2016}. One of the prototype examples is the two-dimensional (2D) time-reversal (TR) TI in which two counterpropagating helical edge electrons manifest, forming a Kramers pair \cite{KaneCL2005,KaneCL2005a,BernevigBA2006,BernevigBA2006a}. The edge state of a 2D TR TI can be described by a helical Luttinger liquid (hLL) \cite{WuC2006,XuC2006}, motivating a substantial number of theoretical studies (see the reviews \cite{HasanMZ2010a,QiXL2011,HasanMZ2011,MaciejkoJ2011,SenthilT2015,DolcettoG2016,RachelS2018,HsuCH2021,WeberB2024} and references therein). The 2D TR TI has a $\mathrm{Z}_2$ topological index, indicating that the odd number of hLLs are topologically nontrivial. In contrast, the even number of hLLs can become Anderson localized without breaking TR symmetry or charge conservation. Theoretically, the topologically nontrivial edge state is ballistic with conductance quantized to $\frac{e^2}{h}$ per edge. The possibility of realizing dissipationless conducting channels is useful for building low-temperature quantum devices.

	\begin{figure}[t!]
		\includegraphics[width=0.4\textwidth]{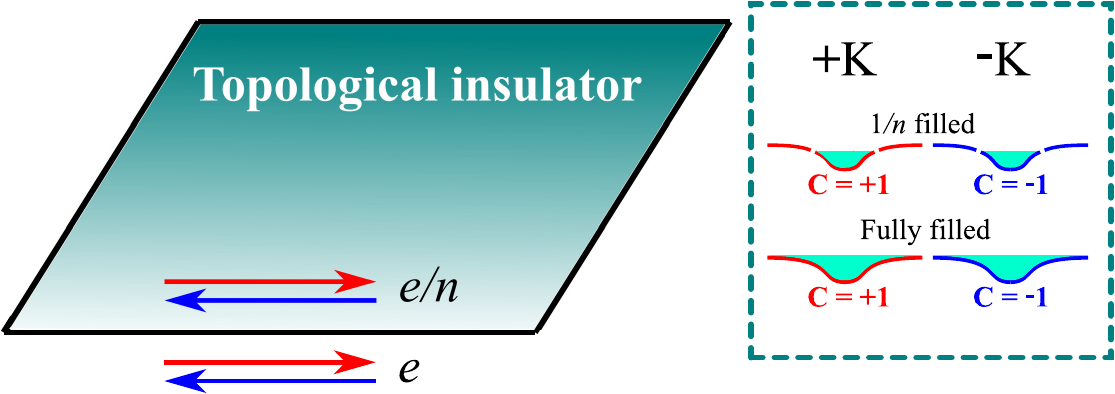}
		\caption{Setup. We consider a composite helical edge state with a regular hLL from fully filled conjugated Chern bands and a (fractional) helical liquid carrying $e/n$ from $1/n$ filled conjugated Chern bands. Right panel: The band structure of a system with two valleys. In the noninteracting limit, the bands in the $+K$ ($-K$) valley have a Chern number $+1$ ($-1$). Note that the spins are locked to the valleys due to the strong Ising spin-orbit coupling in the twisted MoTe$_2$, resulting in spin up in the $+K$ valley and spin down in the $-K$ valley. The lowest bands of both valleys are filled, and the second bands are $1/n$-filled (with $n=1,2,3$). For $n=2,3$, the interaction induces a charge gap, and the system becomes a fractional topological insulator. Left panel: Illustration of the composite helical edge state, consisting of a regular hLL and a helical liquid carrying $e/n$ charge.
		}
		\label{Fig:setup}
	\end{figure}

	After nearly two decades of the birth of 2D TR TI theory \cite{KaneCL2005,KaneCL2005a,BernevigBA2006,BernevigBA2006a}, there have been abundant of candidate materials, e.g., HgTe \cite{KonigM2007,RothA2009}, InAs/GaSb \cite{KnezI2011,DuL2015,LiT2015}, WSe$_2$ \cite{WuS2018,UgedaMM2018}, WTe$_2$ \cite{TangS2017,FeiZ2017}, bismuthine \cite{ReisF2017}, germanene \cite{BampoulisP2023}, AB stacked MoTe$_2$/WSe$_2$ \cite{LiT2021}, and TaIrTe$_4$ \cite{TangJ2024}. However, the transport signature often shows unexpected resistance, i.e., the conductance is less than $\frac{e^2}{h}$ per channel. The fundamental reason for the unexpected resistance is that the carriers move in both directions, allowing for electron backscattering by breaking the protected symmetry \cite{WuC2006,XuC2006}. By contrast, the integer quantum Hall edge states have movers in the same direction, and the perturbations give unimportant random phases to the electrons. The experimental situation has motivated numerous theoretical studies trying to understand the sources and conditions of the conductance reduction \cite{WuC2006,MaciejkoJ2009,TanakaY2011,SchmidtTL2012,LezmyN2012,VayrynenJI2013,VayrynenJI2014,KainarisN2014,ChouYZ2015,VayrynenJI2018}. For example, the helical edge state can become insulating due to spontaneous TR symmetry breaking (TRSB), either through coupling to external magnetic moments \cite{AltshulerBL2013,HsuCH2017,HsuCH2018} or via interaction \cite{WuC2006,XuC2006,KainarisN2014,ChouYZ2018,NovelliP2019,BubisAV2021}. The stability of multiple Kramers pairs has also been explored theoretically \cite{XuC2006,SantosRA2015,SantosRA2016,KagalovskyV2018,SantosRA2019,ChouYZ2019}, suggesting a rich phase diagram due to the interplay between interaction and disorder. The complications due to disorder and interactions potentially hinder the identification of 2D TR TIs, and a cleaner, more controllable material platform is desirable.

	A modern approach to creating new quantum materials is to stack and twist 2D Van der Waal materials, such as graphene \cite{AndreiEY2020,BalentsL2020,PixleyJH2019,AndreiEY2021} and transition metal dichalcogenides \cite{MakKF2022} (e.g., twisted transition metal dichalcogenides homobilayers may realize 2D TR TIs \cite{WuF2019b}). Recently, fractional quantum anomalous Hall effect (without a magnetic field) has been observed in twisted MoTe$_2$ \cite{CaiJ2023a,ZengY2023a,ParkH2023,XuF2023a} and rhombohedral pentalayer graphene aligned with hBN \cite{LuZ2024}. These new experiments open up new directions for realizing unprecedented quantum materials with strong correlation. More surprisingly, a new experiment of twisted MoTe$_2$ at $\theta=2.1^{\circ}$ \cite{KangK2024c} reported a possible fractional TI (FTI) state at hole doping $\nu_{\text{tot}}=3$ ($\nu_{\text{tot}}$ is the number of carriers per moir\'e unit cell) with an edge-state conductance $\frac{3}{2}\frac{e^2}{h}$. They also provide evidence of single, double, and triple helical edge states, suggesting that all three moir\'e bands are characterized by a Chern number $+1$ for the $+K$ valley and by a Chern number $-1$ for the $-K$ valley, forming pairs of conjugated Chern bands. The edge state at $\nu_{\text{tot}}=3$ can be viewed as a composition of a regular helical liquid carrying $e$ (from a pair of fully filled conjugated Chern bands) and a fractionalized channel carrying $e/2$ (from a pair of half-filled conjugated Chern bands). The possibility of realizing an FTI or a fractional quantum spin Hall insulator (with $S_z$ conservation) has motivated several theoretical studies focusing on half-integer FTI states \cite{ZhangYH2024,May-MannJ2024,JianCM2024,VilladiegoIS2024,ZhangYH2024a,ReddyAP2024,XuC2024,ChenF2024}. In particular, an Abelian $Z_4$ topological order \cite{JianCM2024,ZhangYH2024a} is proposed, and the corresponding edge state is the simplest stable helical edge state, described by two counterpropagating charged bosons carrying $e/2$. In this work, we focus on the stable Abelian topological order edge states.
	
	The primary characterization of FTI is based on a vanishing Hall conductance concomitant with a quantized ballistic edge-state conductance, relying on the properties of the edge state. To understand the putative FTI in the twisted MoTe$_2$ experiment \cite{KangK2024c}, it is crucial to study the phase diagram of the composite helical edge state at $\nu_{\text{tot}}=3$, incorporating disorder and interaction. The situation here is more complicated than a single fractionalized channel as the interaction between the regular helical liquid and the fractionalized channel may significantly alter the phase diagram based on a single helical liquid, as shown in the two regular helical liquids setups \cite{SantosRA2015,SantosRA2016,KagalovskyV2018,ChouYZ2019}. Thus, a systematic study of the composite helical edge state incorporating disorder and interaction is needed.

	In this work, we investigate the composite $(1+1/n)$ Abelian helical edge state consisting of a regular hLL and a (fractionalized) helical liquid carrying $e/n$ with $n=1,2,3$. Such a composite $(1+1/n)$ helical edge state arises in a topological insulator as illustrated in Fig.~\ref{Fig:setup}, where a pair of conjugated Chern bands are fully filled and the higher conjugated Chern bands are $1/n$ filled. We develop a general framework for the composite $(1+1/n)$ Abelian helical edge state. The $n=1$ case ($\nu_{\text{tot}}=4$) corresponds to two regular hLLs \cite{XuC2006,SantosRA2015,SantosRA2016,ChouYZ2019}. For $n=2$ ($\nu_{\text{tot}}=3$), we consider a regular hLL and the edge state from an Abelian $Z_4$ FTI \cite{JianCM2024,ZhangYH2024a}. For $n=3$ ($\nu_{\text{tot}}=8/3$), we study a regular hLL and a pair of time-reversal Laughlin $1/3$ edge states \cite{BernevigBA2006}. The results for $n=3$ can be generalized to other odd integers $n>3$. We construct phase diagrams and analyze the conductance for each phase, providing useful criteria for future experimental characterizations. We further investigate the effect of an applied magnetic field and discuss the implications to the twisted MoTe$_2$ experiment \cite{KangK2024c}.

	The rest of the paper is organized as follows: First, we quickly review and discuss the single Abelian (fractionalized) helical liquid in Sec.~\ref{Sec:1hLL}. In Sec.~\ref{Sec:Gen_framework}, we introduce our general framework for the composite $(1+1/n)$ Abelian helical edge, including the bosonization convention, the symmetry-allowed backscattering operators and their scaling dimensions, the stability and the compatibility of perturbations, assumption of an approximate $S_z$ conservation, and the edge-state conductance. We construct the generic phase diagram and compute the edge-state conductance for the $n=1$ case in Sec.~\ref{Sec:n=1}, for the $n=2$ case (which is relevant to the recent twisted MoTe$_2$ \cite{KangK2024c}) in Sec.~\ref{Sec:n=2}, and for the $n=3$ in Sec.~\ref{Sec:n=3}. The effects of a magnetic field on various phases are investigated in Sec.~\ref{Sec:B_field}. We discuss several issues and implications to the experiments in Sec.~\ref{Sec:Discussion}. 
	
	Many technical details are discussed in the appendices. In Appendix~\ref{App:Mapping_SA}, we map the quadratic free boson action to the symmetric and antisymmetric sectors. In Appendix~\ref{App:Bosonization}, we discuss the bosonization of the backscattering operators. In Appendix~\ref{App:Stability}, the stability and the compatibility conditions are studied. In Appendix~\ref{App:excitation}, we investigate the excitations in different insulating edge states. In Appendix~\ref{App:Conductance}, the two-terminal edge-state conductance formalism is constructed and discussed. An alternative conductance derivation is provided in Appendix~\ref{App:Alt_cond}.

	\section{Single helical liquid}\label{Sec:1hLL}
	
	We discuss a standing-alone Abelian (fractional) TI edge state carrying $e/n$ charge, setting up the context for the two helical liquid problems. The $n=1$ case is a regular helical Luttinger liquid arising from a TI edge state \cite{WuC2006,XuC2006,HasanMZ2010a,QiXL2011}. For the $n=2$ case, we consider the $Z_4$ FTI state from conjugated Chern bands \cite{JianCM2024,ZhangYH2024a}. We consider the time-reversal partners of Laughlin $1/3$ FQH states for the $n=3$ \cite{BernevigBA2006}. In the rest of the section, we briefly discuss the edge theory for $n=2$ and $n=3$. Then, the phase diagram is constructed.
	
	\subsection{Abelian $Z_4$ topological order for $n=2$}
	
	The $n=2$ edge state is associated with an FTI at $\nu_{\text{tot}}=1/2+1/2=1$.
	The simplest helical edge state manifests on the boundary of an FTI forming an Abelian $Z_4$ topological order proposed in Refs.~\cite{JianCM2024,ZhangYH2024a}. This state is related to the conjugated composite fermion liquids \cite{ZhangYH2018,ZhangYH2024,Myerson-JainN2023,ShiZD2024} but with a different flux attachment procedure \cite{JianCM2024,ZhangYH2024a}. Alternatively, the same $Z_4$ state can be obtained by condensing product topological orders (i.e., time-reversal partners of FQH at $\nu=1/2$) \cite{JianCM2024}. It is noteworthy that that the $Z_4$ state was also proposed in the contexts of half-filled quantum Hall bilayers \cite{SodemannI2017} and flat TI bands \cite{PotterAC2017}, predating the FTI experiment \cite{KangK2024c}. 
	
	In the Chern-Simons theory for the Abelian $Z_4$ FTI, the K matrix is given by $\hat{K}=\begin{pmatrix} 0 & 4 \\ 4 &0 \end{pmatrix}$. The edge state can be described by an imaginary-time action given by \cite{ZhangYH2024,JianCM2024}
	\begin{align}
		\mathcal{S}_{Z_4}=\frac{4}{4\pi}\int d\tau dx\left[\begin{array}{c} \left(\partial_x\varphi_c\right)\left(i\partial_{\tau}\varphi_s\right)+\left(\partial_x\varphi_s\right)\left(i\partial_{\tau}\varphi_c\right)
		\end{array}
		\right]+\dots,
	\end{align}
	where $\varphi_c$ and $\varphi_s$ correspond to the bosons carrying charge and spin, respectively. We have ignored the kinetic terms in the above expression. In this theory, $e^{-i\varphi_c}$ creates a quasiparticle carrying $(Q,S_z)=(1/2,0)$, and $e^{-i\varphi_s}$ creates a quasiparticle carrying $(Q,S_z)=(0,1/4)$. $Q$ is the charge in the unit of $e$, and $S_z$ is the $z$-component spin. Using $\varphi_R=\varphi_c+\varphi_s$ and $\varphi_L=\varphi_c-\varphi_s$, the edge theory can be expressed by
	\begin{align}
		\label{Eq:S_2}\mathcal{S}^{(n=2)}_{1\text{hLL}}=\frac{2}{4\pi}\int d\tau dx\left[\begin{array}{r}
			\left(\partial_x\varphi_R\right)\left(i\partial_{\tau}\varphi_R\right)\\[2mm]
			-\left(\partial_x\varphi_L\right)\left(i\partial_{\tau}\varphi_L\right)
		\end{array}\right]+\dots,
	\end{align}
	where we have ignored the terms associated with the velocity matrix. The vertex operator $e^{-i\varphi_R}$ creates a right moving excitation with $(Q,S_z)=(1/2,1/4)$. We note that the local single electron excitation is gapped in the edge state of the Abelian $Z_4$ topological order. The minimal charge excitation is $e^{-i2(\varphi_R+\varphi_L)}$  ($Q=2$), and the minimal spin excitation is $e^{-i2(\varphi_R-\varphi_L)}$ ($S_z=1$).
	
	We note that the product topological order $U(1)_8\times U(1)_{-8}$ also gives rise to a pure bosonic edge theory with the same form as Eq.~(\ref{Eq:S_2}). However, the minimal charge is $e/4$ \cite{May-MannJ2024} (rather than $e/2$ in the $Z_4$ theory). Based on the criterion introduced by Levin and Stern \cite{LevinM2009c,LevinM2012,SternA2016}, such an edge state is unstable and can become insulating without breaking TR symmetry or charge $U(1)$ conservation. The Abelian $Z_4$ order edge state \cite{JianCM2024,ZhangYH2024a} is stable and remains ballistic as long as TR symmetry and charge $U(1)$ conservation are intact.
	
	\subsection{Time-reversal partners of Laughlin $1/3$ state for $n=3$}
	
	The FTI associated with $n=3$ can be realized by a product topological order -- time-reversal partners of $\nu=1/3$ FQH Laughlin states \cite{BernevigBA2006,NeupertT2014,KlinovajaJ2014a}. The edge states are made of two counterpropagating charge bosons that carry $e/3$ charges. This product topological order can be described by a Chern-Simons theory with a K matrix $\hat{K}=\begin{pmatrix}
		3 & 0\\ 0 & -3
	\end{pmatrix}$. The edge theory is described by the imaginary-time action as follows:
	\begin{align}
		\mathcal{S}^{(n=3)}_{1\text{hLL}}=\frac{3}{4\pi}\int d\tau dx\left[\begin{array}{r}
			\left(\partial_x\varphi_R\right)\left(i\partial_{\tau}\varphi_R\right)\\[2mm]
			-\left(\partial_x\varphi_L\right)\left(i\partial_{\tau}\varphi_L\right)
		\end{array}\right]+\dots,
	\end{align}
	where we have ignored the velocity matrix terms. The vertex operator $e^{-i\varphi_R}$ creates a quasiparticle with $(Q,S_z)=(1/3,1/6)$; the vertex operator $e^{-i\varphi_L}$ creates a quasiparticle with $(Q,S_z)=(1/3,-1/6)$. The single electron excitation with $S_z=1/2$ ($S_z=-1/2$) can be created by $e^{-i3\varphi_R}$ ($e^{-i3\varphi_L}$).

	\subsection{General edge theory}
	
	The $1/n$ edge state can be described by a chiral Luttinger liquid theory \cite{WenXG1992,ChangAM2003} with the imaginary-time action given by
	\begin{align}
		\mathcal{S}^{(n)}_{1\text{hLL}}=\frac{n}{4\pi}\int d\tau dx \left[\begin{array}{c}
			\left(\partial_x\varphi_{R}\right)\left(i\partial_{\tau}\varphi_{R}+v\partial_x\varphi_R\right)\\[2mm]
			+\left(\partial_x\varphi_{L}\right)\left(-i\partial_{\tau}\varphi_L+v\partial_x\varphi_L\right)\\[2mm]
			-2v'\left(\partial_x\varphi_{R}\right)\left(\partial_x\varphi_{L}\right)
		\end{array}
		\right],
	\end{align}
	where $\varphi_{R}$ ($\varphi_{L}$) is the right (left) moving chiral boson, $v$ is the velocity, and $v'$ encodes the interaction between two chiral movers. $v'>0$ denotes repulsive interaction, and $|v'|<v$ is required for the stability of the bosonic theory. It is customary to define the Luttinger parameter $K=\sqrt{(v-v')/(v+v')}$ to describe the strength of interaction. $K<1$ ($K>1$) indicates a repulsive (attractive) $v'$. The chiral bosons obey the following commutation relations
	\begin{align}
		[\partial_x\varphi_{R}(x),\varphi_{R}(x')]=&+2\pi in^{-1}\delta(x-x'),\\
		[\partial_x\varphi_{L}(x),\varphi_{L}(x')]=&-2\pi in^{-1}\delta(x-x').
	\end{align} 
	The densities are defined by $\rho_{R}=\frac{1}{2\pi}\partial_x\varphi_{R}$ and $\rho_{L}=-\frac{1}{2\pi}\partial_x\varphi_{L}$, and the charge $e$ annihilation operators are $R\sim e^{in\varphi_R}$ and $L\sim e^{in\varphi_L}$. We can show that $[\frac{1}{2\pi}\partial_x\varphi_R,e^{in\varphi_R}]=-e^{in\varphi_R}\delta(x-x')$ and $[\frac{-1}{2\pi}\partial_x\varphi_L,e^{in\varphi_L}]=-e^{in\varphi_L}\delta(x-x')$, consistent with $[R^{\dagger}R(x),R(x')]=-R(x)\delta(x-x')$ and $[L^{\dagger}L(x),L(x')]=-L(x)\delta(x-x')$.
	
	Finally, the theory obeys a spinful TR symmetry: The theory is invariant under $R\rightarrow L$, $L\rightarrow -R$, and $i\rightarrow -i$. Equivalently, the TR operations can be expressed by $\varphi_R\rightarrow-\varphi_L$, $\varphi_L\rightarrow-\varphi_R+\pi/n$, and $i\rightarrow -i$.
	
	\subsection{Phase diagram}
	
	The primary TR symmetric disorder to the edge state is the long-wavelength density fluctuation (i.e., $R^{\dagger}R+L^{\dagger}L$), which does not induce any instability by itself. Meanwhile, the single-electron backscattering (e.g., $L^{\dagger}R+R^{\dagger}L$) violates the TR symmetry. The leading TR symmetric backscattering perturbation is an umklapp two-particle backscattering \cite{XuC2006,WuC2006}, e.g., $:(R^{\dagger}L)^2:+$H.c., where $:A:$ denotes the normal ordering of $A$, corresponding to $\cos(2n\varphi_R-2n\varphi_L)$ in the bosonization. In a disorder-free system, such an interaction only has an effect near the commensurate fillings (i.e., $2k_F\approx G$, where $G$ is the reciprocal lattice vector). The same is true for other higher-order umklapp backscattering. 
	
	The combination of TR symmetric disorder and the backscattering interaction can relax the umklapp condition, resulting in random local commensuration. Technically, the situation can be viewed as a commensurate backscattering interaction with a spatially varying coupling constant. (There are minor differences in the results \cite{ChouYZ2018}, but these are unimportant for this work.) For a sufficiently strong repulsive interaction ($K<\frac{3}{8n}$), a TRSB localized insulating state can be realized \cite{WuC2006,XuC2006,KainarisN2014,ChouYZ2018}. In summary, there are two possible phases: A ballistic phase for $K>\frac{3}{8n}$ and a TRSB insulating state for $K<\frac{3}{8n}$. As we discuss later in this work, the phase diagram of the composite $(1+1/n)$ helical edge is much richer and complicated.
	
	\section{General framework for composite ($1+1/n$) Abelian helical edge}\label{Sec:Gen_framework}
	
	In this section, we introduce a general framework for studying the composite $(1+1/n)$ Abelian helical edge state. We assume that the 2D bulk is unaffected in the presence of disorder and interactions in the edge states. We further focus only on long edges and the low temperatures, where field theory predictions apply. First, we discuss the chiral boson theory and define the rescaled variables for mapping to two regular hLLs problems. Then, we discuss various perturbations and the role of the Rashba spin-orbit coupling. Finally, we study the edge-state conductance.
	
	\subsection{Model: Free bosons}
	
	To model the composite $(1+1/n)$ Abelian helical edge state, we employ the chiral boson description \cite{WenXG1992,ChangAM2003}. In the imaginary-time path integral, the action of an edge state is described by
	\begin{align}\label{Eq:S_n}
		\mathcal{S}^{(n)}_{2\text{hLL}}=\frac{1}{4\pi}\int dxd\tau\left(\partial_x\Phi^T\right)\left[\hat{K}\left(i\partial_{\tau}\Phi\right)+\hat{V}\left(\partial_x\Phi\right)\right],
	\end{align}
	where $\Phi^T=[\varphi_{1R},\varphi_{1L},\varphi_{2R},\varphi_{2L}]$, $\varphi_{aR}$ ($\varphi_{aL}$) is the chiral boson field for the $a$th channel ($a=1,2$) moving toward the $+x$ ($-x$) direction,
	\begin{align}
		\label{Eq:K_matrix}\hat{K}=&\left[\begin{array}{cccc}
			1 & 0 & 0 & 0\\
			0 & -1 & 0 & 0\\
			0 & 0 & n & 0\\
			0 & 0 & 0 & -n
		\end{array}
		\right],\\
		\label{Eq:V_matrix}\hat{V}=&\left[\begin{array}{cccc}
			v_1 & -v_1' & \sqrt{n}u & -\sqrt{n}u'\\
			-v_1' & v_1 & -\sqrt{n}u' & \sqrt{n}u\\
			\sqrt{n}u & -\sqrt{n}u' & nv_2 & -nv_2'\\
			-\sqrt{n}u' & \sqrt{n}u & -nv_2' & nv_2
		\end{array}
		\right].
	\end{align} 
	In the expression of $\hat{V}$, there are 6 independent parameters \cite{XuC2006} encoding the edge-state velocities ($v_1$ and $v_2$) and interactions ($v_1'$, $v_2'$, $u$, and $u'$). As pointed out in \cite{XuC2006}, three independent ``boost'' parameters are sufficient to determine the scaling dimensions of operators. Since $\hat{V}$ must be positive definite, all the eigenvalues of $\hat{V}$ must be positive, providing constraints to the parameters. The factor $n$ and $\sqrt{n}$ in $\hat{V}$ are due to the normalization convention. The minus signs in $\hat{V}$ ensure that the repulsive (attractive) interaction corresponds to positive (negative) $v_1'$, $v_2'$, and $u'$ in our convention. The chiral bosons obey the following commutation relation:
	\begin{subequations}\label{Eq:Comm}
		\begin{align}
			[\partial_x\varphi_{1\eta}(x),\varphi_{1\eta'}(x')]=&2\pi i\zeta_{\eta} \delta_{\eta,\eta'}\delta(x-x'),\\
			[\partial_x\varphi_{2\eta}(x),\varphi_{2\eta'}(x')]=&2\pi i\zeta_{\eta} n^{-1} \delta_{\eta,\eta'}\delta(x-x'),\\
			[\partial_x\varphi_{1\eta}(x),\varphi_{2\eta'}(x')]=&[\partial_x\varphi_{2\eta}(x),\varphi_{1\eta'}(x')]=0,
		\end{align}
	\end{subequations}
	where $\eta=R,L$, $\zeta_R=+1$, and $\zeta_L=-1$.
	The density operator is defined by $\rho_{a\eta}=\frac{\zeta_{\eta}}{2\pi}\partial_x\varphi_{a\eta}$.
	Thus, the charge $e$ annihilation operators are described by $R_1(x)\sim e^{i\varphi_{aR}(x)}$, $L_1(x)\sim e^{i\varphi_{aL}(x)}$, $R_2(x)\sim e^{in\varphi_{aR}(x)}$, and $L_2(x)\sim e^{in\varphi_{aL}(x)}$. One can confirm the above relations by checking $[R_j^{\dagger}R_j(x),R_j(x')]=-\delta(x-x')R_j(x)$, $[L_j^{\dagger}L_j(x),L_j(x')]=-\delta(x-x')L_j(x)$.
	
	In the quantum spin Hall limit, the $z$-component of spin $S_z$ is a good quantum number, and the $R_a$ ($L_a$) corresponds to a $S_z=1/2$ ($S_z=-1/2$) state.
	Thus, the spinful TR operation is given by $R_a\rightarrow L_a$, $L_a\rightarrow -R_a$, and $i\rightarrow -i$. In the absence of $S_z$ conservation (e.g., the presence of Rashba spin-orbit coupling), the edge state satisfies the same TR operation, but the spin texture becomes momentum-dependent \cite{RodA2015,XieHY2016}. Any TR symmetric perturbation to the edge state must be invariant under the TR operation.

	\subsection{Rescaling of variables}
	
	The two helical liquids problem given by Eq.~(\ref{Eq:S_n}) can be mapped to the $n=1$ case corresponding to two regular hLLs. We introduce the rescaled chiral boson fields $\tilde{\Phi}=[\tilde{\varphi}_{1R},\tilde{\varphi}_{1L},\tilde{\varphi}_{2R},\tilde{\varphi}_{2L}]^T$, defined by $\tilde\varphi_{1R}=\varphi_{1R}$, $\tilde\varphi_{1L}=\varphi_{1L}$, $\tilde\varphi_{2R}=\sqrt{n}\varphi_{2R}$, and $\tilde\varphi_{2L}=\sqrt{n}\varphi_{2L}$. The action of the edge state becomes
	\begin{align}\label{Eq:S_n'}
		\mathcal{S}^{(n)}_{2\text{hLL}}=\frac{1}{4\pi}\int dxd\tau\left(\partial_x\tilde\Phi^T\right)\left[\hat{\tilde{K}}\left(i\partial_{\tau}\tilde\Phi\right)+\hat{\tilde{V}}\left(\partial_x\tilde\Phi\right)\right],
	\end{align}
	where
	\begin{align}
		\label{Eq:K_V_matrix'}\hat{\tilde{K}}=&\left[\begin{array}{cccc}
			1 & 0 & 0 & 0\\
			0 & -1 & 0 & 0\\
			0 & 0 & 1 & 0\\
			0 & 0 & 0 & -1
		\end{array}
		\right],\hat{\tilde{V}}=&\left[\begin{array}{cccc}
			v_1 & -v_1' & u & -u'\\
			-v_1' & v_1 & -u' & u\\
			u & -u' & v_2 & -v_2'\\
			-u' & u & -v_2' & v_2
		\end{array}
		\right].
	\end{align}
	The $\hat{\tilde{K}}$ and $\hat{\tilde{V}}$ are independent of $n$ with the rescaled fields in $\tilde{\Phi}$. The chiral bosonic fields can be rewritten by $\tilde{\varphi}_{aR}=\phi_a+ \theta_a$ and $\tilde{\varphi}_{aL}=\phi_a-\theta_a$, where $\phi_a$ and $\theta_a$ correspond to the phase and density bosonic fields, respectively. The charge $e$ fermionic excitations can be easily expressed by $\theta$'s and $\phi$'s. 
	\begin{subequations}\label{Eq:RL_boson}
		\begin{align}
			R_{1}\sim& e^{i\varphi_{1R}}=e^{i\tilde{\varphi}_{1R}}=e^{i(\phi_1+\theta_1)},\\
			L_{1}\sim& e^{i\varphi_{1L}}=e^{i\tilde{\varphi}_{1L}}=e^{i(\phi_1-\theta_1)},\\
			R_{2}\sim& e^{in\varphi_{2R}}=e^{i\sqrt{n}\tilde{\varphi}_{2R}}=e^{i\sqrt{n}(\phi_2+\theta_2)},\\
			L_{2}\sim& e^{in\varphi_{2L}}=e^{i\sqrt{n}\tilde{\varphi}_{2L}}=e^{i\sqrt{n}(\phi_2-\theta_2)}.
		\end{align}
	\end{subequations}
	Thus, the physical charge operators are given by $\rho_1=\frac{e}{\pi}\partial_x\theta_1$ and $\rho_2=\frac{e}{ \sqrt{n}\pi}\partial_x\theta_2$, where $e<0$ is the electric charge.

	In the two hLLs problem, it is more natural to describe the problem using the collective variables, $\phi_{\pm}=\left(\phi_1\pm\phi_2\right)/\sqrt{2}$ and $\theta_{\pm}=\left(\theta_1\pm\theta_2\right)/\sqrt{2}$. We simplify the problem by assuming $v_1=v_2=v$ and $v_1'=v_2'=v'$. The action given by Eq.~(\ref{Eq:S_n'}) can be expressed by (see Appendix~\ref{App:Mapping_SA} for detailed derivations)
	\begin{align}
		\nonumber\mathcal{S}^{(n)}_{2hLL}\rightarrow&\int d\tau dx\frac{1}{\pi}\left[\left(i\partial_\tau\phi_+\right)\left(\partial_x\theta_+\right)+\left(i\partial_\tau\phi_-\right)\left(\partial_x\theta_-\right)\right]\\
		\nonumber&+\int d\tau dx\frac{v_+}{2\pi}\left[K_+\left(\partial_x\phi_+\right)^2\frac{1}{K_+}\left(\partial_x\theta_+\right)^2\right]\\
		\label{Eq:S_+-_decoupled}&+\int d\tau dx\frac{v_-}{2\pi}\left[K_-\left(\partial_x\phi_-\right)^2\frac{1}{K_-}\left(\partial_x\theta_-\right)^2\right],
	\end{align}
	where $v_+$ and $K_+$ ($v_-$ and $K_-$) correspond to the velocity and Luttinger parameter for the channel-symmetric (channel-antisymmetric) sector. 
	In our convention, the Luttinger liquid parameter $K_{\pm}<1$ ($K_{\pm}>1$) means effective repulsive (attractive) interaction in the $\pm$ sector. Note that $K_->1$ can be achieved by pure repulsive microscopic interactions as long as the interchannel repulsion is stronger than the intrachannel repulsion as discussed in Appendix~\ref{App:Mapping_SA}.

	\subsection{Symmetry allowed perturbation}\label{Sec:gen_fr:OPs}
	
	Now, we discuss TR symmetric perturbations in the composite $(1+1/n)$ Abelian helical edge state. The analysis is based on symmetry rather than any specific model. We enumerate all the symmetric-allowed single-particle and two-particle backscattering processes most relevant in the renormalization group (RG) sense. The pure forward-scattering perturbations either contribute to a random phase factor or renormalize the Luttinger parameters. These effects do not induce qualitative different results in our analysis. We briefly review the conditions for relevant RG flows of uniformed and disordered perturbations. Then, the operators associated with different TR symmetric backscattering processes are discussed.

	In a translation-invariant edge state, the perturbation can be described by
	\begin{align}\label{Eq:S_X_uni}
		\mathcal{S}_{\mathcal{X},\text{uniform}}=U_{\mathcal{X}}\int d\tau dx\left[e^{i\delta_{\mathcal{X}}x}\mathcal{O}_{\mathcal{X}}(\tau,x)+\text{H.c.}\right],
	\end{align}
	where $\mathcal{X}$ specifying the type of perturbation, $U_{\mathcal{X}}$ is the strength of the perturbation, $\mathcal{O}_{\mathcal{X}}$ is an operator, and $\delta_{\mathcal{X}}$ is the total wavevector (modulo to the lattice wavevector). This perturbation is generically irrelevant unless momentum conservation or umklapp condition is satisfied, i.e., $|\delta_{\mathcal{X}}|$ is smaller than a threshold associated with the commensurate-incommensurate transition \cite{PokrovskyVL1979}. Thus, the $\mathcal{S}_{\mathcal{X},\text{uniform}}$ can be ignored for general nonzero $\delta_{\mathcal{X}}$. The generic edge state is not translation invariant, and disorder prevails in every sample boundary. Thus, we consider a disordered edge state with position-dependent white noise Gaussian random potentials. For each operator $\mathcal{O}_{\mathcal{X}}$, the perturbation can be described by
	\begin{align}\label{Eq:S_X_dis}
		\mathcal{S}_{\mathcal{X},\text{dis}}=\int d\tau dx\left[\xi_{\mathcal{X}}(x)\mathcal{O}_{\mathcal{X}}(\tau,x)+\text{H.c.}\right],
	\end{align}
	where $\xi_{\mathcal{X}}(x)$ is a zero-mean Gaussian random variable obeying $\overline{\xi_{\mathcal{X}}(x)}=0$,
	$\overline{\xi_{\mathcal{X}}(x)\xi^*_{\mathcal{X}}(x')}=W_{\mathcal{X}}\delta(x-x')$, and $W_{\mathcal{X}}$ is the variance of disorder. ($\overline{A}$ means disorder average of $A$.)
	
	In the RG analysis, the relevance of $\mathcal{X}$ perturbation is determined by the scaling dimension $\Delta_{\mathcal{X}}$ and the Kosterlitz-Thouless RG flow. For a uniform perturbation with commensuration [i.e., Eq.~(\ref{Eq:S_X_uni}) with $\delta_{\mathcal{X}}\approx0$], the RG flow for $U_{\mathcal{X}}$ is given by
	\begin{align}
		\frac{dU_{\mathcal{X}}}{dl}=\left(2-\Delta_{\mathcal{X}}\right)U_{\mathcal{X}}.
	\end{align}
	Thus, $U_{\mathcal{X}}$ becomes relevant when $\delta_{\mathcal{X}}\approx0$ and $\Delta_{\mathcal{X}}<2$. For a disorder perturbation [i.e., Eq.~(\ref{Eq:S_X_dis})], the RG equation for $W_\mathcal{X}$ (after disorder average) is given by 
	\begin{align}
		\frac{dW_{\mathcal{X}}}{dl}=\left(3-2\Delta_{\mathcal{X}}\right)W_{\mathcal{X}}.
	\end{align}
	$W_\mathcal{X}$ becomes relevant when $\Delta_{\mathcal{X}}<3/2$ \cite{GiamarchiT1988}. While a relevant disordered perturbation relaxes momentum conservation, a smaller scaling dimension (compared to the uniformed commensurate perturbation) is generally required, indicating a stronger interaction is needed in general. The scaling dimension $\Delta_{\mathcal{X}}$ can be computed using the framework developed in \cite{MooreJE1998,XuC2006}, and three independent ``boost'' parameters are needed \cite{XuC2006}. In this work, we evaluate the scaling dimensions assuming Eq.~(\ref{Eq:S_+-_decoupled}), corresponding to only two independent parameters, $K_+$ and $K_-$. Our general results do not rely on this standard approximation \cite{KlesseR2000,ChouYZ2015,ChouYZ2019} for two Luttinger liquids problems.
	
	\begin{figure}[t!]
		\includegraphics[width=0.3\textwidth]{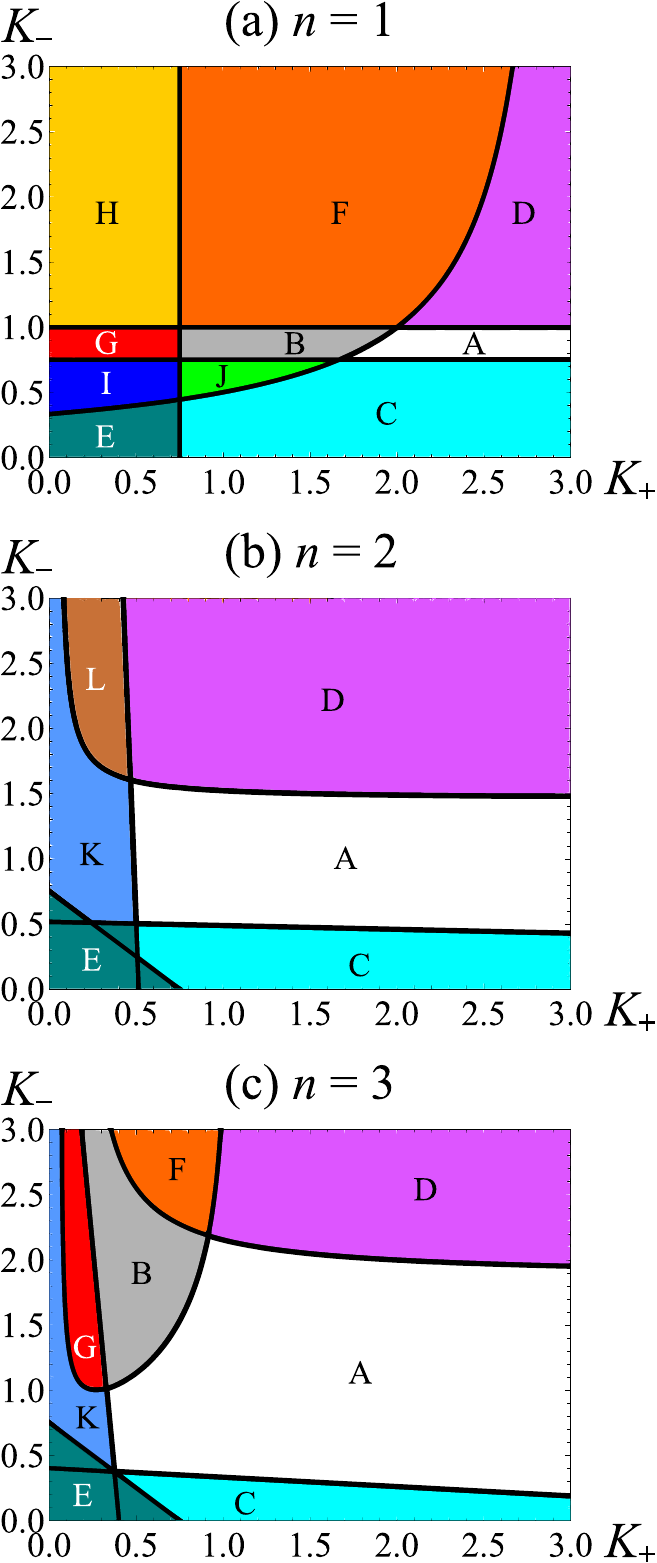}
		\caption{Regions of relevant perturbations. Each line indicates $\Delta_{\chi}=3/2$ for $\mathcal{O}_{\chi}\neq \mathcal{O}_J$ and $\Delta_{J}=2$. We use different colors to mark the regions with different relevant perturbations. Region A: There is no relevant perturbation. Region B: $\mathcal{O}_{M,1}$ and $\mathcal{O}_{M,1}$ are relevant. Region C: Only $\mathcal{O}_-$ is relevant. Region D: Only $\mathcal{O}_J$ is relevant. Region E: At least two out of $\mathcal{O}_{+}$, $\mathcal{O}_{-}$, $\mathcal{O}_{2p,1}$ are relevant. Region F: $\mathcal{O}_{M,1}$, $\mathcal{O}_{M,2}$, and $\mathcal{O}_{J}$ are relevant. Region G: $\mathcal{O}_{M,1}$, $\mathcal{O}_{M,2}$, and $\mathcal{O}_{+}$ are relevant. Region H: $\mathcal{O}_{M,1}$, $\mathcal{O}_{M,2}$, $\mathcal{O}_+$, and $\mathcal{O}_{J}$ are relevant. Region I: $\mathcal{O}_{M,1}$ and $\mathcal{O}_{M,2}$ are relevant. At least two out of $\mathcal{O}_{+}$, $\mathcal{O}_{-}$, $\mathcal{O}_{2p,1}$ are also relevant. Region J: $\mathcal{O}_{M,1}$, $\mathcal{O}_{M,2}$, and $\mathcal{O}_-$ are relevant. Region K: Only $\mathcal{O}_+$ is relevant. L: $\mathcal{O}_+$ and $\mathcal{O}_J$ are relevant. The perturbations $\mathcal{O}_{M,1}$ and $\mathcal{O}_{M,2}$ are absent for the $n=2$ theory studied in this work, so the line $\Delta_M=3/2$ is absent in (b). 
		}
		\label{Fig:Scaling}
	\end{figure}

	Now, we discuss various TR symmetric operators.
	First, the TR symmetric single-electron backscattering bilinears \cite{XuC2006} can be constructed as follows
	\begin{subequations}\label{Eq:O_M}
		\begin{align}
			\label{Eq:O_M1}\mathcal{O}_{M,1}=&L_2^{\dagger}R_1-R_2^{\dagger}L_1+\text{H.c.}\\
			\label{Eq:O_M2}\mathcal{O}_{M,2}=&iL_2^{\dagger}R_1+iR_2^{\dagger}L_1+\text{H.c.}.
		\end{align}
	\end{subequations}
	These two operators describe backscattering charge $e$ between two channels, corresponding to $\Delta S_z=1$ processes. The TR symmetry forbids the single-electron backscattering within a channel. The scaling dimension is given by 
	\begin{align}
		\Delta_{M}=\frac{(1+\sqrt{n})^2}{8}\left(K_++K_-^{-1}\right)+\frac{(1-\sqrt{n})^2}{8}\left(K_-+K_+^{-1}\right).
	\end{align}
	The expression for $\Delta_{M}$ is reduced to $\frac{1}{2}(K_++K_-^{-1})$ for $n=1$. For $n>1$, the minimal $\Delta_{M}=(n-1)/2$ is obtained with $K_+=\frac{\sqrt{n}-1}{\sqrt{n}+1}$ and $K_-=\frac{\sqrt{n}+1}{\sqrt{n}-1}$. This term is absent for the $Z_4$ FTI edge state \cite{JianCM2024,ZhangYH2024a}, which is the $n=2$ case in this work.

	The leading TR symmetric backscattering interactions correspond to the four-fermion operators given by
	\begin{align}
		\label{Eq:O_+}\mathcal{O}_{+}=&L_1^{\dagger}R_1L_2^{\dagger}R_2+\text{H.c.},\\
		\label{Eq:O_-}\mathcal{O}_{-}=&L_1^{\dagger}R_1R_2^{\dagger}L_2+\text{H.c.}.
	\end{align} 
	For $n=1$, $\mathcal{O}_{+}$ and $\mathcal{O}_{-}$ can be viewed as the interaction in the channel-symmetric and channel-antisymmetric sectors \cite{ChouYZ2019}, respectively. We will show that $\mathcal{O}_{+}$ ($\mathcal{O}_{-}$) generally induces a negative (positive) drag among two channels. $\mathcal{O}_{+}$ is a $\Delta S_z=2$ process, while $\mathcal{O}_-$ corresponds to $\Delta S_z=0$. The scaling dimensions are given by
	\begin{align}
		\Delta_+=&\frac{(1+\sqrt{n})^2}{2}K_++\frac{(1-\sqrt{n})^2}{2}K_-,\\
		\Delta_-=&\frac{(1+\sqrt{n})^2}{2}K_-+\frac{(1-\sqrt{n})^2}{2}K_+.
	\end{align}
	For $n=1$, $\Delta_+$ ($\Delta_-$) is minimized to 0 when $K_+=0$ ($K_-=0$). For $n>1$, $\Delta_{+}$ and $\Delta_-$ are minimized to zero when $K_+=K_-=0$.
	
	The two-particle backscattering operator within the channel $a$ is given by \cite{WuC2006,XuC2006}
	\begin{align}
		\label{Eq:O_2p}\mathcal{O}_{2p,a}=:\left(L^{\dagger}_aR_a\right)^2:+\text{H.c.},
	\end{align}
	where $:\mathcal{A}:$ denotes the normal order of $\mathcal{A}$. The intra-channel two-particle backscattering corresponds to $\Delta S_z=2$. The corresponding scaling dimension is given by
	\begin{align}
		\Delta_{2p,1}=2(K_++K_-),\,\,\,\Delta_{2p,2}=2n(K_++K_-).
	\end{align}
	The minimal $\Delta_{2p,a}=0$ can be obtained by setting $K_{+}=K_-=0$.

	Finally, a Josephson coupling between two channels is also allowed and is described by
	\begin{align}
		\label{Eq:O_J}\mathcal{O}_J=R^{\dagger}_1L^{\dagger}_1L_2R_2+\text{H.c.}.
	\end{align}
	The Josephson coupling is a $\Delta S_z=0$ process. The scaling dimension is given by
	\begin{align}
		\Delta_J=\frac{(1+\sqrt{n})^2}{2}K_-^{-1}+\frac{(1-\sqrt{n})^2}{2}K_+^{-1}.
	\end{align}
	For $n=1$, $\Delta_{J}$ is minimized to zero when $K_-\rightarrow \infty$.
	For $n>1$, the minimal $\Delta_{J}=0$ can be obtained by setting $K_{+}\rightarrow\infty$ and $K_{-}\rightarrow\infty$.

	Among all the operators discussed above, $\mathcal{O}_J$ is the only perturbation associated with a zero wavevector. As a result, a uniform Josephson coupling interaction can manifest without fine tuning the Fermi wavevector. We generically assume disordered perturbations given by Eq.~(\ref{Eq:S_X_dis}) for other perturbations. Therefore, the phase diagram is determine by $(\Delta_J-2)$ and $(\Delta_{\mathcal{X}}-3/2)$ for $\mathcal{O}_{\mathcal{X}}\neq\mathcal{O}_J$. We label the regions based on relevant operators in Fig.~\ref{Fig:Scaling}. The results suggest that the phase diagram is complicated and nonuniversal. Later, we will show that universal phase diagrams can be derived by assuming an approximate $S_z$ conservation on the edge state.

	\subsection{Stability and compatibility criteria}
	
	The analysis in Sec.~\ref{Sec:gen_fr:OPs} provides the conditions for which the perturbations are relevant under the weak coupling RG, indicating the instability of the free boson action given by Eq.~(\ref{Eq:K_V_matrix'}) [or Eq.~(\ref{Eq:S_+-_decoupled})]. When the perturbation associated with $\mathcal{O}_{\mathcal{X}}$ becomes relevant, the (weak-coupling) RG flows tend to develop a large $U_{\mathcal{X}}$ or $W_{\mathcal{X}}$ (depending on the type of perturbation). In such a situation, the low-energy manifold is significantly modified, and a strong coupling analysis is needed. Here, we apply the strong coupling analysis introduced by Haldane \cite{HaldaneFDM1995} to the composite $(1+1/n)$ helical edge problem. The original bosonic variables without rescaling [which are introduced in Eq.~(\ref{Eq:S_n})] are used for the following discussion.

	For a vertex operator representing a physical backscattering process (including single-particle and interaction), it can be expressed by $\exp[im^T\Phi(\tau,x)]$, where $m$ is a four-component vector of integers. The corresponding scaling dimension $\Delta(m)$ is bounded by \cite{XuC2006,MooreJE2002}
	\begin{align}
		\label{Eq:D_m}\Delta(m)\ge \left|\frac{m^T\hat{K}^{-1}m}{2}\right|\equiv \mathcal{D}(m).
	\end{align}
	When $\mathcal{D}(m)=0$ (the null-vector condition), the combination of the modes given by $m^T\Phi$ can be removed from the low-energy theory in the strong coupling fixed point \cite{HaldaneFDM1995,MooreJE1998,XuC2006,NeupertT2011,NeupertT2014}. When multiple vertex operators fulfill the null-vector condition, the vertex operators might not be compatible with each other due to nontrivial commutation relations. For two vectors $m_1$ and $m_2$, the corresponding vertex operators can independently remove the low-energy modes if $m_1^T\hat{K}^{-1}m_2=0$ and $\mathcal{D}(m_1)=\mathcal{D}(m_2)=0$ \cite{HaldaneFDM1995,NeupertT2011,NeupertT2014}. Note that the results here are universal and independent of the velocity matrix $\hat{V}$. These criteria provide useful information for the low-energy theory in the strong coupling limit.

	For a physical backscattering process with $\mathcal{D}(m)\neq 0$, the corresponding strong coupling theory is much subtler. This situation is similar to the disordered edge state of an FQH insulator at $\nu=2/3$ \cite{KaneCL1994,KaneCL1995,ProtopopovIV2017}, where the fixed point (a.k.a. the Kane-Fisher-Polchinski fixed point) manifests a charge mode and a neutral mode. Later, we show that the $n=3$ case can be mapped to two copies of the $\nu=2/3$ FQH edges.
	
	We examine the lower bound of the scaling dimension for each perturbation discussed in Sec.~\ref{Sec:gen_fr:OPs}. We find that all the perturbations except $\mathcal{O}_{M,a}$ [given by Eq.~(\ref{Eq:O_M})] with $n>1$ fulfill the null vector condition. However, multiple incompatible perturbations can be simultaneously relevant under the weak coupling RG flows in several parameter regions as plotted in Fig.~\ref{Fig:Scaling}. We summarize the compatibility of different vertex operators in Appendix~\ref{App:Stability}. These results can also be understood intuitively using the scaling dimensions derived in Sec.~\ref{Sec:gen_fr:OPs} with $K_+$ and $K_-$. The null-vector condition corresponds to a zero scaling dimension at some $(K_+,K_-)$. When two distinct perturbations (that obey the null-vector condition) are compatible, there is a special point of $(K_+,K_-)$ such that the two scaling dimensions are simultaneously zero. The scaling dimensions in Sec.~\ref{Sec:gen_fr:OPs} are consistent with the general results summarized in Appendix~\ref{App:Stability}, suggesting that the properties of the operators can be sufficiently described by our simplified scheme with $K_+$ and $K_-$.

	\subsection{Phase diagram with approximate $S_z$ conservation}\label{Sec:Ge_fr:Sz}
	
	Many TR TI systems also possess an approximate (but unnecessary) $S_z$ symmetry, which makes spin-flipping perturbations parametrically small. Here, we impose an approximate $S_z$ conservation by assuming a hierarchy of perturbation strengths based on the change of $S_z$, $\Delta S_z$. Specifically, we treat the $\Delta S_z =0$ terms as dominant perturbations, the $\Delta S_z=1$ terms as the subleading perturbations, and the $\Delta S_z=2$ terms as the weakest perturbations. The hierarchy of perturbation strength is useful for situations in which multiple incompatible terms are relevant under weak coupling RG. The perturbation with the smallest $\Delta S_z$ dominates the low-energy phase.
	
	With the assumptions mentioned above, we identify six possible phases: a phase of free bosons without any relevant backscattering interactions, a TRSB localized insulator, the $\alpha$ phase with dominant $\mathcal{O}_{M,1}$ and $\mathcal{O}_{M,2}$ perturbations, the $\beta$ phase with a dominant $\mathcal{O}_-$ perturbation, the $\gamma$ phase with a dominant $\mathcal{O}_J$ perturbation, and the $\delta$ phase with the dominant $\mathcal{O}_+$ perturbation. In addition, the universal generic phase diagrams are constructed in Figs.~\ref{Fig:PD_n1}, \ref{Fig:PD_n2}, and \ref{Fig:PD_n3}. We briefly discuss the general properties of each phase in the following.
	
	The TRSB localized insulator here is a two-channel generalization of the single-channel TRSB localized insulator discussed in Sec.~\ref{Sec:1hLL}. It has a minimal charge excitation of $e/(2n)$, and the ground state can be viewed as an XY spin glass \cite{ChouYZ2018}. The nature of the $\alpha$ phase depends on $n$. The $\alpha$ phase is a topologically trivial Anderson insulator with TR symmetry in the $n=1$ case. However, for $n=3$, the $\alpha$ phase is always metallic, and the transport coefficient depends on the microscopic details, analogous to the $\nu=2/3$ FQH effect \cite{KaneCL1994,KaneCL1995,ProtopopovIV2017}. The $\alpha$ phase is absent for $n=2$ as the single-electron excitation is not allowed in the $Z_4$ edge state \cite{JianCM2024,ZhangYH2024a}. The $\beta$ phase can be viewed as a positive perfect Coulomb drag phase \cite{NazarovYV1998,KlesseR2000}. The $\gamma$ phase can be viewed as a phase-locked Josephson coupled phase. The $\beta$ and $\gamma$ phases have exactly the same ground state manifold, but the Luttinger parameters are different, facilitating different instabilities. The $\delta$ phase can be viewed as a negative perfect Coulomb drag phase \cite{FuruyaSC2015,ChouYZ2019,ChouYZ2023f}. The phases mentioned above also have properties that are specifically $n$-dependent, and we discuss these results in depth later.

	\subsection{Edge-state conductance}\label{Sec:gen_fr:cond}
	
	In addition to identifying the low-energy phases, we are interested in studying the measurable quantities. For 2D TR TIs, the characterization is primarily based on transport: Absence of Hall conductance and quantized edge-state conductance. Thus, it is crucial to develop a conductance theory for each phase discussed in this work. For interacting 1D systems connected to two external electron reservoirs, the dc transport does not depend on the forward-scattering interaction (e.g., the Luttinger liquid interaction), and the conductance can be computed using a Landauer setup \cite{MaslovDL1995,PonomarenkoVV1995,SafiI1995}. There are two scenarios that the edge-state conductance is modified from the noninteracting value: (a) the perturbation satisfies the null-vector condition \cite{HaldaneFDM1995,MooreJE1998,XuC2006,NeupertT2011,NeupertT2014}, indicating that a combination of the modes can be removed from the low-energy theory. (b) the edge states are subjected to perturbations that do not satisfy the null-vector condition. Both scenarios are discussed in the following.

	We derive the edge-state conductance for scenario (a), generalizing the Oreg-Sela-Stern formalism \cite{OregY2014,CornfeldE2015,ShavitG2019} to our case with fractionalized charge channels. Remarkable, we find that the $\beta$ and $\gamma$ phases give an edge-state conductance $G=(1+1/n)\frac{e^2}{h}$, the same as the free bosons phase, despite that the edge state is strongly interacting. The results can be understood by the perfect positive drag between the two channels. Meanwhile, the $\delta$ phase gives an edge-state conductance $G=(1-1/n)\frac{e^2}{h}$, showing an example of unusual quantized conductance not directly associated with the filling factor $\nu_{\text{tot}}$. $G=(1-1/n)\frac{e^2}{h}$ is a manifestation of the perfect negative drag of the $\delta$ phase. The derivation of the results can be found in Appendix.~\ref{App:Conductance}. 
	
	The scenario (b) is realized when $\mathcal{O}_{M,a}$ operator is the dominant perturbation for $n=3$. 
	In particular, such a case is mathematically pertinent to two copies of $\nu=2/3$ FQH edge states \cite{KaneCL1994,KaneCL1995,ProtopopovIV2017} that form time-reversal partners. When the two copies of $\nu=2/3$ edge states are decoupled [i.e., $v_1'=v_2'=u=0$ in $\hat{V}$ given by Eq.~(\ref{Eq:V_matrix})], we can apply the existing results for the $\nu=2/3$ FQH edge states \cite{ProtopopovIV2017} to our case with $n=3$. A detailed discussion can be found in Sec.~\ref{Sec:n=3}.

	\section{Composite helical edge at $n=1$: Two ordinary helical Luttinger liquids}\label{Sec:n=1}
	
	\begin{figure}[t!]
		\includegraphics[width=0.35\textwidth]{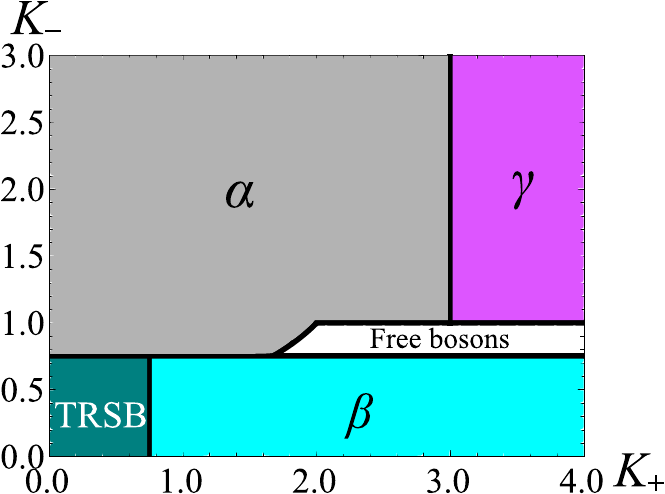}
		\caption{Generic phase diagram for $n=1$. We consider a weak Rashba spin-orbit coupling such that the total $S_z$ is approximately conserved. The TRSB and $\alpha$ phases are insulating. The free bosons, $\beta$, and $\gamma$ phases give conductance $G=2\frac{e^2}{h}$.}
		\label{Fig:PD_n1}
	\end{figure}

	The $n=1$ case corresponds to two ordinary hLLs arising from non-fractionalized TIs, related to the so-called double helical edge states in several experiments \cite{KangK2024c,KangK2024a}. This problem has been studied in several previous work, focusing on the stability of edge states \cite{XuC2006,SantosRA2015,SantosRA2016,KagalovskyV2018} and applications to Coulomb drag systems \cite{ChouYZ2015,KainarisN2017,ChouYZ2019}. Here, we revisit the problem and consider all the leading symmetry-allowed perturbations in a disordered edge state. The analysis and results of the $n=1$ case also provide useful guidance for the situation with $n> 1$.
	
	\subsection{Phase diagram}
	
	The two hLLs problem can be expressed in terms of channel-symmetric and channel-antisymmetric sectors \cite{ChouYZ2019}, analogous to the charge and spin in a 1D spinful system, respectively. In this work, we consider a simplified case in which the symmetric and antisymmetric sectors are decoupled in the quadratic free boson action, described by Eq.~(\ref{Eq:S_+-_decoupled}).
	Using Eq.~(\ref{Eq:S_+-_decoupled}) and the perturbations discussed in Sec.~\ref{Sec:gen_fr:OPs}, we obtain the conditions of relevant perturbations and summarize the results in Fig.~\ref{Fig:Scaling}(a), which can be viewed as the phase diagram without enforcing $S_z$ conservation. However, the resulting phase diagram is nonuniversal and depends on the microscopic detail. Alternatively, we assume an approximate $S_z$ symmetry (i.e., small Rashba spin-orbit coupling) because of the experimental situations \cite{KangK2024c,KangK2024a}. (See Sec.~\ref{Sec:Ge_fr:Sz} for a discussion.) This auxiliary assumption produces a hierarchy of perturbation strengths based on $\Delta S_z$, the change of $S_z$ during the process. In particular, $\mathcal{O}_-$ and $\mathcal{O}_J$ are the strongest perturbations with $\Delta S_z=0$; the $\mathcal{O}_{M,1}$ and $\mathcal{O}_{M,2}$ are the subleading perturbation with $\Delta S_z=1$; the $\mathcal{O}_{+}$, $\mathcal{O}_{2p,1}$, and $\mathcal{O}_{2p,2}$ are treated as the weakest perturbation with $\Delta S_z=2$. The generic phase diagram with the approximate $S_z$ conservation is plotted in Fig.~\ref{Fig:PD_n1}. We discuss the phases in the following.
	
	The dominant perturbations $\mathcal{O}_-$ and $\mathcal{O}_J$ lead to two distinct channel-symmetric fluid phases named phases $\beta$ and $\gamma$, respectively. The $\beta$ phase is related to the perfect positive Coulomb drag between two channels \cite{NarozhnyBN2016,NazarovYV1998,KlesseR2000,ChouYZ2015,ChouYZ2019,ChouYZ2023f}, while the $\gamma$ phase can be viewed as phase-locked Josephson coupled channels. Besides the Luttinger parameters, the primary differences between these two phases are in the localized/gapped channel-antisymmetric sectors: A charge-density-wave-like order takes place in the $\beta$ phase, and a superconducting-like order is present in the $\gamma$ phase. 
	The subleading perturbation $\mathcal{O}_{M,a}$ is compatible with $\mathcal{O}_J$, suggesting that two perturbations can coexist in the strong coupling limit. The dominant $\mathcal{O}_J$ pins the value of $\phi_-$ and then makes the scaling dimension $\Delta_{M,a}\rightarrow K_+/2$, enhancing the relevant region of $\mathcal{O}_{M,a}$. As a result, the Anderson localization (the $\alpha$ phase), induced by $\mathcal{O}_{M,a}$, takes place for a wide range of parameters (including the noninteracting point $K_+=K_-=1$) as shown in Fig.~\ref{Fig:PD_n1}. Meanwhile, $\mathcal{O}_{M,a}$ is suppressed for $K_-<3/4$ because the dominant perturbation $\mathcal{O}_-$ is not compatible with the subleading $\mathcal{O}_{M,a}$. The suppression of $\mathcal{O}_{M,a}$ is a manifestation of the ``interaction-protected'' topological edge states \cite{SantosRA2015,SantosRA2016}. For $K_+<3/4$ and $K_-<3/4$, a TRSB localized insulator \cite{ChouYZ2018,ChouYZ2019} is realized as the $\mathcal{O}_+$ and $\mathcal{O}_-$ are relevant and mutually compatible. The TRSB insulator is not adiabatically connected to the Anderson insulator \cite{ChouYZ2018,ChouYZ2019}, and the charge excitation is $e/2$. There is also a small region coined free bosons indicating that no perturbation is relevant.  
	
	In principle, the $\delta$ phase with dominant $\mathcal{O}_+$ can exist in the two hLLs problem, exhibiting perfect negative Coulomb drag \cite{ChouYZ2019,DuL2021} of two hLLs. In our case, such a phase is preempted by the $\alpha$ phase, i.e., Anderson localization. The $\delta$ phase might occur when two hLLs have negligible electron tunneling, i.e., neligible $\mathcal{O}_{M,a}$. In the phase diagram in Fig.~\ref{Fig:PD_n1}, we assume a hierarchical structure of the perturbations based on $\Delta S_z$, and $\delta$ phase is absent. 
	
	We briefly comment on the differences between the TR symmetric Anderson and TRSB localized insulators. The former is adiabatically connected to a noninteracting limit, and the charge excitation is the regular unit electron charge $e$. The latter case is very different from the noninteracting Anderson insulator, and the charge excitation is $e/2$, half electric charge \cite{QiXL2008}. The half-charge excitations are related to the kinks and anti-kinks in the bosonized perturbation operators \cite{ChouYZ2018,ChouYZ2019} (see Appendix~\ref{App:excitation}). The ground state of a TRSB localized insulator possesses spatially varying expectation values of the $S_x$ and $S_y$ order parameters \cite{ChouYZ2018,ChouYZ2019,ChouYZ2021b}, reminiscent of an XY spin glass state.

	\subsection{Conductance}
	
	We discuss the conductance for each phase discussed above. First, all the perturbations for $n=1$ satisfy Haldane's null-vector condition, indicating that the low-energy modes associated with the perturbation can be removed from the low-energy theory. The conductance for each phase in Fig.~\ref{Fig:PD_n1} can be computed using the Oreg-Sela-Stern formalism \cite{OregY2014,CornfeldE2015,ShavitG2019} (see Appendix~\ref{App:Conductance}) or by a direct analysis based on the localized/gapped modes (see Appendix~\ref{App:Alt_cond}). The $\alpha$ and TRSB phases are localized with conductance $G\propto \exp(-2L/\xi_{\text{loc}})$ for $L\gg \xi_{\text{loc}}$, where $L$ is the edge length and $\xi_{\text{loc}}$ is the localization length. Meanwhile, the free bosons, $\beta$, and $\gamma$ phases have the same ballistic conductance $2\frac{e^2}{h}$. Our results show that the ballistic transport can happen in the interacting phases driven by $\Delta S_z=0$ perturbations (i.e., $\beta$ and $\gamma$ phases) in addition to the free bosons phase.

	\section{Composite helical edge at $n=2$: A regular helical Luttinger liquid and an $e/2$ helical liquid}\label{Sec:n=2}
	
	\begin{figure}[t!]
		\includegraphics[width=0.35\textwidth]{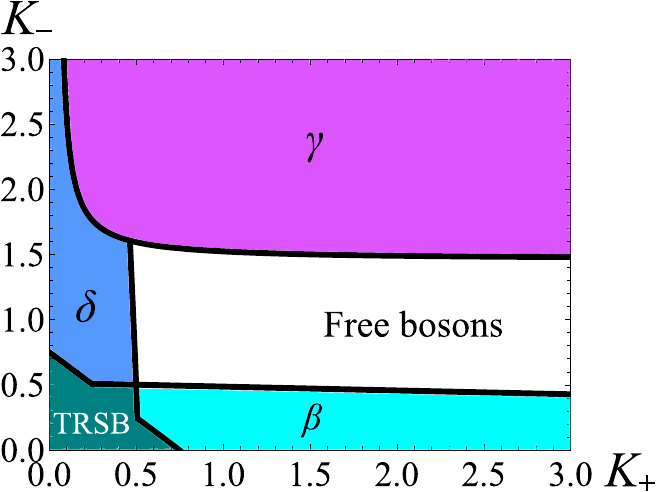}
		\caption{Generic phase diagram for $n=2$. We consider a weak Rashba spin-orbit coupling such that the total $S_z$ is approximately conserved. We focus on the Abelian helical liquid arising from the Abelian $Z_4$ topological order \cite{ZhangYH2024a,JianCM2024} for channel 2. In this case, the $\alpha$ phase is absent because single-electron excitation is gapped in channel 2. The conductance of each phase is summarized as follows: $G=\frac{3}{2}\frac{e^2}{h}$ for the free boson, $\beta$, and $\gamma$ phases; $G=\frac{1}{2}\frac{e^2}{h}$ for the $\delta$ phase; $G\approx0$ for the TRSB localized phase with a sufficiently long edge length.
		}
		\label{Fig:PD_n2}
	\end{figure}	
	
	In this section, we study the composite edge state with a regular hLL and a fractionalized state carrying $e/2$. This composite state is related to the twisted MoTe$_2$ experiment \cite{KangK2024c} showing a possible FTI at  $\nu_{\text{tot}}=3$. We consider the simplest stable Abelian FTI edge state -- the Abelian $Z_4$ topological order edge state arising from half-filled conjugated Chern bands \cite{JianCM2024,ZhangYH2024a}. Other possible FTI edge states, such as the product of $\nu=1/2$ FQH states, have been discussed in Ref.~\cite{May-MannJ2024}. We discuss the phase diagram and the two-terminal conductance for the composite $(1+1/2)$ Abelian helical edge.

	\subsection{Phase diagram}
	
	First, we construct a phase diagram of the $n=2$ case consisting of a regular hLL and an Abelian $Z_4$ helical liquid carrying $e/2$. In Fig.~\ref{Fig:Scaling}(b), the relevant regions of various operators are shown, suggesting a complicated phase diagram without $S_z$ conservation. Note that the inter-channel single-electron backscattering ($\mathcal{O}_{M,a}$) is absent as the electron excitation is gapped in the Abelian $Z_4$ topological order edge state \cite{JianCM2024,ZhangYH2024a}. With the assumption of an approximate $S_z$ conservation (see Sec.~\ref{Sec:Ge_fr:Sz}), a generic phase diagram with a weak Rashba spin-orbit coupling is plotted in Fig.~\ref{Fig:PD_n2}. The phase diagram consists of the free bosons phase without any relevant perturbation, the TRSB localized insulator, the $\beta$ phase showing a perfect positive Coulomb drag between the two channels, the $\gamma$ phase showing phase-locked Josephson coupled channels, and the $\delta$ phase showing a perfect negative Coulomb drag between the two channels. Similar to the $n=1$ case,  the $\beta$ and $\gamma$ phases have the same low-energy bosonic modes, manifesting a positive current drag, $j_2=j_1/2$, where $j_a$ is the current in the $a$th channel (see Appendix~\ref{App:Alt_cond}). The differences between the $\beta$ and $\gamma$ phases are the gapping/localizing mechanisms and the Luttinger parameters. The $\delta$ phase is also partially gapped/localized, and the low-energy modes feature a perfect negative drag, $j_2=-j_1/2$ (see Appendix~\ref{App:Alt_cond}). The TRSB localized insulator is qualitatively similar to the $n=1$ case. The difference is that the minimal charge excitation in the second channel is $e/4$ (see Appendix~\ref{App:excitation}). Again, the $\alpha$ phase is absent for $n=2$ because the $Z_4$ state has gapped electron excitation in the second channel \cite{JianCM2024,ZhangYH2024a}.
	
	The absence of the $\alpha$ phase widens the region of the free bosons phase. Meanwhile, the $\beta$ and TRSB phases have smaller regions than the $n=1$ case because of the scaling dimensions of operators are generically larger compared to $n=1$. In the absence of the forward scattering interactions (i.e., $K_+=K_-=1$), the composite helical edge is in the free bosons phase, suggesting that ballistic edge states are likely to be observed in an FTI supporting a $(1+1/2)$ composite helical edge state. Next, we turn to the two-terminal conductance for all the above mentioned phases.

	\subsection{Conductance}
	
	Now, we discuss the conductance of the composite $(1+1/2)$ helical edge state. All the perturbations in this case obey Haldane's null-vector criteria \cite{HaldaneFDM1995}. Thus, we can compute the conductance using the methods mentioned in Appendices~\ref{App:Conductance} and \ref{App:Alt_cond}. We summarize the results in the following. The free bosons phase gives the ballistic conductance, $\frac{3}{2}\frac{e^2}{h}$. The $\beta$ and $\gamma$ phases form perfect positive current drag, and the conductance is also $\frac{3}{2}\frac{e^2}{h}$. The $\delta$ phase features a perfect negative drag, yielding a conductance $\frac{1}{2}\frac{e^2}{h}$. The TRSB localized insulator has a conductance $\propto \exp(-2L/\xi_{\text{loc}})$ for $L\gg \xi_{\text{loc}}$, where $L$ is the edge length and $\xi_{\text{loc}}$ is the localization length. The ballistic conductance, $\frac{3}{2}\frac{e^2}{h}$, can be found for a wide range of parameters (in the free bosons, $\beta$, and $\gamma$ phases) in Fig.~\ref{Fig:PD_n2}. The result of the $\delta$ phase suggests an unusual quantized conductance $\frac{1}{2}\frac{e^2}{h}$ in the composite $(1+1/2)$ helical edge, indicating a potential complication for experimental verification of the bulk FTI.

	\section{Composite helical edge at $n=3$: A regular helical Luttinger liquid and an $e/3$ helical liquid}\label{Sec:n=3}

	\begin{figure}[t!]
		\includegraphics[width=0.35\textwidth]{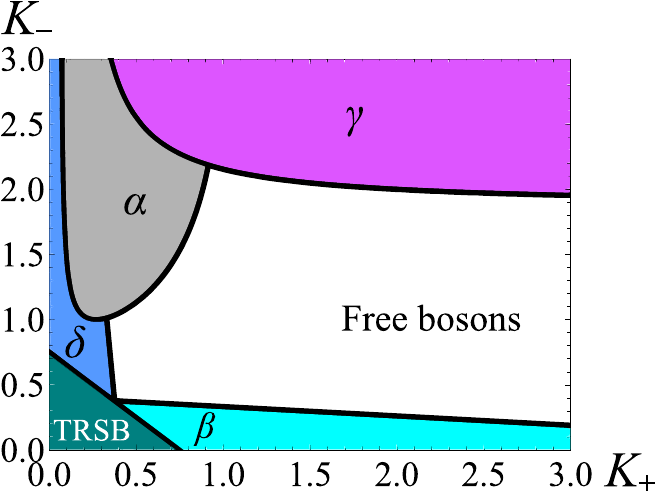}
		\caption{Generic phase diagram for $n=3$. We consider a weak Rashba spin-orbit coupling such that the total $S_z$ is approximately conserved. The conductance of the $\alpha$ phase may depend on microscopic details, similar to the $\nu=2/3$ FQH effect \cite{KaneCL1994,KaneCL1995,ProtopopovIV2017}. The conductance of other phases is summarized as follows: $G=\frac{4}{3}\frac{e^2}{h}$ for the free boson, $\beta$, and $\gamma$ phases;  $G=\frac{2}{3}\frac{e^2}{h}$ for the $\delta$ phase; $G\rightarrow0$ for the TRSB phase with a sufficiently long edge length.
		}
		\label{Fig:PD_n3}
	\end{figure}	
	
	In this section, we discuss a composite helical edge state made of a regular hLL and an Abelian helical liquid carrying an $e/3$ charge. Such an Abelian helical liquid can arise from the FTI state with time-reversal partners of $\nu=1/3$ Laughlin states \cite{BernevigBA2006,NeupertT2014,KlinovajaJ2014a}, but it has not been suggested experimentally. We theoretically investigate the phase diagram of the composite $(1+1/3)$ helical edge and the two-terminal conductance in the rest of the section.

	\subsection{Phase diagram}
	
	First, we examine the scaling dimensions and identify the relevant perturbations as shown in Fig.~\ref{Fig:Scaling}(c), which contain eight distinguishable regions. With the assumption discussed in Sec.~\ref{Sec:Ge_fr:Sz}, a generic phase diagram with a weak Rashba spin-orbit coupling is constructed in Fig.~\ref{Fig:PD_n3}. We find the free bosons phase without relevant perturbation, the TRSB localized insulator, the $\alpha$ phase dictated by $\mathcal{O}_{M,1}$ and $\mathcal{O}_{M,2}$, the $\beta$ phase showing a perfect positive Coulomb drag between the two channels, and the $\gamma$ phase showing phase-locked Josephson coupled channels, and the $\delta$ phase showing a perfect negative Coulomb drag between the two channels. Similar to the $n=1$ and $n=2$ cases, the $\beta$ and $\gamma$ phases are the perfect positive drag phases with the same low-energy bosonic modes, but the gapping/localizing mechanisms and Luttinger parameters are different. The $\delta$ phase is a perfect negative drag phase. The TRSB localized insulator contains the minimal $e/2$ and $e/6$ charge excitations in channels 1 and 2, respectively. (See Appendix~\ref{App:excitation} for a discussion of the charge excitations.) The $\alpha$ phase here is a TR symmetric disorder-dominated metal (different from the Anderson localization for the $n=1$ case). We discuss the $\alpha$ phase next.
	
	In the $n=3$ case, $\mathcal{O}_{M,1}$ and $\mathcal{O}_{M,2}$ do not satisfy the null-vector criteria \cite{HaldaneFDM1995} (the minimal scaling dimension$\mathcal{\Delta}_{M,a}=1$), and the $\alpha$ phase remains metallic. The $\alpha$ phase is related to the disordered FQH edge at $\nu=2/3$ \cite{KaneCL1994,KaneCL1995,ProtopopovIV2017}. In the absence of $v_1'$, $v_2'$, and $u$ in the velocity matrix $\hat{V}$ [Eq.~(\ref{Eq:V_matrix})], the composite $(1+1/3)$ helical edge can be viewed as two decoupled time-reversal partners of $\nu=2/3$ FQH edge states. Specifically, $\varphi_{1R}$ and $\varphi_{2L}$ can be viewed as a $\nu=2/3$ FQH edge state, and $\varphi_{1L}$ and $\varphi_{2R}$ forms another $\nu=2/3$ FQH edge state that is the time-reversal partner of the former. Under such a circumstance, the RG flow for each copy leads to the Kane-Fisher-Polchinski fixed point \cite{KaneCL1994,MooreJE2002}, in which the $\nu=2/3$ FQH edge state is decomposed into decoupled charge and neutral modes \cite{KaneCL1994,KaneCL1995,ProtopopovIV2017}. We expect that the RG flow in the general situation [i.e., with a general $\hat{V}$ in Eq.~(\ref{Eq:V_matrix})] also lead to the Kane-Fisher-Polchinski fixed point corresponding to the minimal scaling dimension $\Delta_{M,a}=1$ \cite{KaneCL1994}. In our phase diagram with $K_+$ and $K_-$, this fixed point is at $K_+=(3-2\sqrt{2})/2$ and $K_-=(3+2\sqrt{2})/2$, which is on the phase boundary separating the $\alpha$ and $\gamma$ phases. For the systems with initial ultraviolet parameters away from the fixed point but still in the $\alpha$ phase, the fixed point can be achieved under the RG flow as long as $\mathcal{O}_{M,1}$ and $\mathcal{O}_{M,2}$ are the dominant perturbations.
	
	The $n=3$ phase diagram is more complicated than the previous two cases. In the absence of the forward scattering interactions (i.e., $K_+=K_-=1$), the composite helical edge is in the free bosons phase, suggesting that the ballistic edge states are likely to be observed in an FTI supporting a composite $(1+1/3)$ Abelian helical edge state. Next, we discuss the two-terminal conductance as a direct experimental probe for the phases uncovered here.
	
	\subsection{Conductance}
	
	The conductance for each phase is discussed here. First, for the perturbation satisfying Haldane's null-vector criteria \cite{HaldaneFDM1995}, we can obtain the conductance using the methods mentioned in Appendices~\ref{App:Conductance} and \ref{App:Alt_cond}. The free bosons, $\beta$, and $\gamma$ phases give $\frac{4}{3}\frac{e^2}{h}$, suggesting ballistic conduction. The $\delta$ phase gives $\frac{2}{3}\frac{e^2}{h}$, indicating a negative drag between two channels. The TRSB localized phase gives an exponentially decaying conductance $\propto \exp(-2L/\xi_{\text{loc}})$ for $L\gg \xi_{\text{loc}}$, where $L$ is the edge length and $\xi_{\text{loc}}$ is the localization length. The conductance in the $\alpha$ phase is more complicated as $\mathcal{O}_{M,1}$ and $\mathcal{O}_{M,2}$ do not satisfy the null-vector criteria. Meanwhile, we can use the existing results in the $\nu=2/3$ FQH edge states to infer the conductance for the $\alpha$ phase.
	
	First, the $\alpha$ phase can be described by the Kane-Fisher-Polchinski fixed point. Based on Ref.~\cite{ProtopopovIV2017}, the conductance at the Kane-Fisher-Polchinski fixed point is in the mesoscopic fluctuation regime with a nonuniversal conductance. Note that this conclusion is different from the results based on Kubo formula, giving a $\frac{2}{3}\frac{e^2}{h}$ edge-state conductance \cite{KaneCL1994}. It is known that the Kubo formula may not predict the experimentally relevant conductance as the coupling to the leads is essential for the two-terminal transport \cite{MaslovDL1995,PonomarenkoVV1995,SafiI1995}. Thus, we expect that the conductance of the $\alpha$ phase is nonuniversal depending on the microscopic details. An interesting implication from Ref.~\cite{ProtopopovIV2017} is that the edge state conductance might saturate to $\frac{2}{3}\frac{e^2}{h}$ in an inelastic regime with irrelevant $\mathcal{O}_{M,1}$ and $\mathcal{O}_{M,2}$. Note that the $\mathcal{O}_{M,1}$ and $\mathcal{O}_{M,2}$ are necessarily much stronger than other perturbations. Therefore, the inelastic condition is only possible in a small region of the free bosons phase sufficiently close to the phase boundary between the $\alpha$ phase. A careful analysis of the conductance of the $\alpha$ phase is useful but beyond the scope of this work. 
	
	Unlike the previous two cases, there is no sharp constraint for the allowed values of the edge-state conductance because of the presence of the $\alpha$ phase. We generally expect that the ballistic conductance can be found experimentally as the free bosons, $\beta$, and $\gamma$ phases all give ballistic conduction. Moreover, the region with ballistic conduction is the largest among the three cases discussed in this work.  
	
	\section{Magnetic field response in Composite helical edges}\label{Sec:B_field}
	
	In this section, we study the magnetic field response in the composite Abelian helical edge states, providing additional signatures based on our theory. We first discuss the field-induced localization of helical edges due to TR symmetry breaking. Then, the effects of a magnetic field on different phases are investigated.
	
	\subsection{Field-induced localization}
	
	We study the effects of an external magnetic field on the composite helical edge states. First, we assume that the (approximate) spin quantization axis of the TI is along the $z$ direction. At the linear order, the external magnetic field gives rise to a coupling Hamiltonian given by \cite{QiXL2008,MaciejkoJ2010,ChouYZ2018}
	\begin{align}
		\hat{H}_B=-t_xB_xS_x-t_yB_yS_y-t_zB_zS_z,
	\end{align}
	where $B_{\mu}$ is the $\mu$-component of the magnetic field, $t_{\mu}$ is the model-dependent coupling constant, and the TR odd spin operators
	\begin{align}
		S_z\sim&\sum_j\int dx\left[R_j^{\dagger}R_j-L^{\dagger}_jL_j\right],\\
		S_x\sim&\sum_j\int dx \left[e^{-i2k_{F,j}x}R^{\dagger}_jL_j+e^{i2k_{F,j}x}L^{\dagger}_jR_j\right],\\
		S_y\sim& \sum_j\int dx\left[-ie^{-i2k_{F,j}x}R^{\dagger}_jL_j+ie^{i2k_{F,j}x}L^{\dagger}_jR_j\right].
	\end{align}
	The $t_z$ term can be viewed as a vector potential that does not induce backscattering in the edge state. Meanwhile, the $t_x$ and $t_y$ terms cause intrachannel single-electron backscattering. We emphasize that the magnetic field must be sufficiently small so that the bulk topology is unaffected.
	
	In the absence of disorder, $t_x$ and $t_y$ can open up a gap at the Dirac point (i.e., $2k_F\approx 0$), and the gap is proportional to $B_{\parallel}=\sqrt{B_x^2+B_y^2}$. For the generic disordered edge states, the $t_x$ and $t_y$ terms become spatially fluctuating and can be mapped to the Giarmachi-Schulz model \cite{GiamarchiT1988,ChouYZ2018}. The weak-coupling RG predicts localization of the first channel for $K_++K_-<3$ and localization of the second channel for $K_++K_-<3/n$. The difference between the two channels is due to the vertex operator expression as discussed in Appendix~\ref{App:Bosonization}. The field-induced localized state has a minimal $e$ charge excitation in the first channel and a minimal $e/n$ charge excitation in the second channel. Note that this TR odd localized state is different from the TRSB localized insulator discussed previously. Based the RG analysis, we obtain the field-dependent localization length $\xi_{B,1}\propto B_{\parallel}^{-2/(3-K_+-K_-)}$ for the first channel and $\xi_{B,2}\propto B_{\parallel}^{-2/(3-nK_+-nK_-)}$ for the second channel. These quantities can be analyzed in a transport experiment with a tunable magnetic field \cite{BubisAV2021}.
	
	The results suggest a possible field-driven phase in the $n>1$ cases where the first channel is insulating but the second channel remains ballistic for $3/n<K_++K_-<3$, giving rise to a conductance $\frac{1}{n}\frac{e^2}{h}$. For a sufficiently strong $B_{\parallel}$, the second channel also becomes localized due to the nature of the sine-Gordon RG flows. Generally, the field-induced localization lengths in the two channels ($\xi_{B,1}$ and $\xi_{B,2}$) are quite different, and a conductance $\frac{1}{n}\frac{e^2}{h}$ is expected for $\xi_{B,1}\ll L\ll \xi_{B,2}$, where $L$ is the edge length. Thus, we anticipate that the magnetic field can cause two transitions in the conductance for $n>1$: $G=(1+1/n)\frac{e^2}{h}\rightarrow \frac{1}{n}\frac{e^2}{h}\rightarrow 0$. 
	
	\subsection{Magnetic field as a probe for different phases}
	
	Now, we discuss the interplay between field-induced localization and other phases. The field-induced localization is not compatible with the $\alpha$ or $\gamma$ phases. Thus, we expect that the field-induced localization happens when the magnetic field is larger than some finite threshold value. For the $\alpha$ phase with $n=1$, the magnetic field drives a localization-localization transition between the inter-channel TR symmetric Anderson localization and the intra-channel field-induced localization. Meanwhile, the field-induced localization is compatible with all the instability associated with only $\theta$'s, such as the $\beta$, $\delta$, and TRSB phases. We note that the $\beta$ and $\delta$ phases become fully insulating when one of the channels is localized due to the applied in-plane magnetic field.
	
	Based on the results discussed above, one can use the magnetic field to characterize various phases uncovered in this work. First, we discuss how to distinguish the free bosons, $\beta$, and $\gamma$ phases, which all give ballistic conductance. The magnetic field generically induces localization in the two channels, characterized by localization lengths $\xi_{B,1}$ (for channel 1) and $\xi_{B,2}$ (for channel 2). The free bosons phase becomes insulating as long as the localization lengths $\xi_{B,1}\ll L$ and $\xi_{B,2}\ll L$, where $L$ is the edge length. For $n> 1$, $\xi_{B,1}\ll\xi_{B,2}$ is possible. As a result, applying a magnetic field to the free bosons phase can induce an intermediate phase with a quantized conductance $G=\frac{1}{n}\frac{e^2}{h}$ for $\xi_{B,1}\ll L\ll\xi_{B,2}$. A fully insulating state is expected for a sufficiently large magnetic field. Since the $\beta$ phase is compatible with the field-induced localization, we expect the conductance rapidly approaches zero (i.e., $\xi_{1,B}\ll L$) in the presence of a small magnetic field. Meanwhile, the $\gamma$ phase is incompatible with the field-induced localization. As a result, the conductance is resilient to the magnetic field, and a finite magnetic field is needed to cause field-induced localization even in the limit $L\rightarrow \infty$.
	
	The magnetic field can also distinguish different localized phases. Applying an in-plane magnetic field to the interaction-driven TRSB localized insulator does not cause a significant change in the conductance as discussed in Ref.~\cite{ChouYZ2018}. Applying an in-plane magnetic field to the $n=1$ $\alpha$ phase can drive a transition. Since the $\alpha$ phase is incompatible with the field-induced localization, the localization length of the TR symmetric state, $\xi_{\text{TR}}$, monotonically increases (i.e., localization is weakened) in the presence of an in-plane magnetic field. For a sufficiently large in-plane magnetic field, $\xi_{B,1}$ and $\xi_{B,2}$  become smaller than $\xi_{\text{TR}}$, and the transport is dominated by the field-induced localization lengths, $\xi_{B,1}$ and $\xi_{B,2}$. Thus, we expect a nonmonotonic conductance as a function of the magnetic field in an edge state with a finite $L$. The maximal conductance (minimal resistance) corresponds to the transition point.

	\section{Discussion}\label{Sec:Discussion}
	
	We study the phase diagrams and the edge-state conductance of the composite $(1+1/n)$ helical edge states with $n=1,2,3$, focusing on the interacting two helical liquids arising from an Abelian (F)TI. First, we develop a systematic framework based on bosonization that maps the original problem into the two interacting hLLs with modified vertex operator perturbations. We provide the conditions for ballistic transport in the composite Abelian helical edge state and further uncover the existence of novel interaction-driven phases, such as the TRSB localized insulator, two distinct perfect positive drag phases (the $\beta$ and $\gamma$ phases), and the perfect negative drag phase (the $\delta$ phase). The single-electron backscattering between two helical liquids also leads to a trivial Anderson localized phase for $n=1$ and a TR symmetric metal described by the Kane-Fisher-Polchinski fixed point \cite{KaneCL1994,KaneCL1995a,ProtopopovIV2017} for $n=3$. We construct the generic phase diagrams (with weak Rashba spin-orbit couplings), compute the edge-state conductance, and discuss using an applied in-plane magnetic field to distinguish different phases with the same conductance. Our work uncovers several new phases that cannot be inferred from FQH edges and establishes an unprecedented systematic theory for the composite helical edge states arising from Abelian FTIs.
	
	In this work, we simplify the velocity matrix [Eq.~(\ref{Eq:V_matrix})] such that the quadratic free bosons action can be described by decoupled channel-symmetric and channel-antisymmetric Luttinger liquids. With this standard simplification for two Luttinger liquids \cite{KlesseR2000,ChouYZ2015,ChouYZ2019}, the scaling dimensions of the operators can be computed straightforwardly. The resulting phase diagrams are constructed by two independent parameters, $K_+$ and $K_-$, indicating the interactions in the associated sectors. Without simplifying the velocity matrix [Eq.~(\ref{Eq:V_matrix})], three independent boost parameters are required for the scaling dimension calculations \cite{XuC2006}. However, constructing phase diagrams with these abstract boost parameters is obscure and without physical intuitions. Notably, our results with $K_+$ and $K_-$ correctly capture the general ``topological'' properties of the operators (i.e., the stability and compatibility conditions discussed in Appendix~\ref{App:Stability}), showing the sufficiency in describing the interplay between different backscattering interactions. Using the general framework \cite{XuC2006} with three boost parameters likely changes some quantitative features, e.g., the phase boundary. Connecting the composite $(1+1/n)$ Abelian helical edge problem to two regular hLLs is an essential technical advance and establishes an intuitive way to study the interacting disordered phase diagrams.
	
	The composite helical edge with $n=2$ is related to an Abelian $Z_4$ topological order \cite{JianCM2024,ZhangYH2024a} arising from half-filled conjugated Chern bands. There are other possible TR topological orders from the conjugated half-filled Chern bands. e.g., product topological order from time-reversal partners of the $U(1)_8$ order, Moore-Read state \cite{MooreG1991}, and other Pfaffian states. These states belong to unstable FTIs (i.e., the edge state can become insulating without breaking TR symmetry) according to the Levin-Stern criterion \cite{LevinM2009c,LevinM2012,SternA2016} as the minimal charge is $e/4$ (rather than $e/2$ in our $Z_4$ state). The disordered interacting edge states of these product topological orders have been studied in Ref.~\cite{May-MannJ2024}. On the contrary, we focus on the edge state of a stable Abelian $Z_4$ FTI \cite{JianCM2024,ZhangYH2024a}, which is not a product topological order.

	To provide useful experimental signatures, we study the two-terminal edge-state conductance for each phase. An unexpected result is an unusual quantized edge-state conductance $G=(1-1/n)\frac{e^2}{h}$ arising from the perfect negative drag phase (the $\delta$ phase) for $n>1$. This conductance is not directly associated with the filling factor $\nu_{\text{tot}}=2+2/n$, indicating the rich, complicated nature of the composite Abelian helical edge states. Since the $\delta$ phase is dictated by the $\mathcal{O}_+$ perturbation with $\Delta S_z=2$, this phase is more likely to be observed in a system with a sufficiently strong Rashba spin-orbit coupling (i.e., without $S_z$ conservation).
	
	Besides the two-terminal conductance, we also investigate the in-plane magnetic field response. In the free bosons phase with $n>1$, the localization lengths in channels 1 and 2 are generically different. Remarkably, we predict a partially localized phase for $n>1$ with localized channel 1 and delocalized channel 2 in the presence of a finite in-plane magnetic field, yielding $G=\frac{1}{n}\frac{e^2}{h}$. The field-induced quantized edge-state conductance can be examined experimentally, providing an important signature of the composite $(1+1/n)$ Abelian helical edge state studied in this work. Addition, the in-plane magnetic field can be used to distinguish the different phases with the same edge-state conductance, suggesting a useful way to characterize the phases discussed in this work.

	Now, we discuss the implications of our theory to the existing FTI experiment \cite{KangK2024c}. In the twisted MoTe$_2$ experiment \cite{KangK2024c}, the non-local measurement suggests an edge-state conductance $G=\frac{3}{2}\frac{e^2}{h}$ at $\nu_{\text{tot}}=3$ concomitant with a nearly vanishing Hall conductance. A possible explanation is a realization of a TR FTI at $\nu_{\text{tot}}=3$. Based on our theory with $n=2$ (Abelian $Z_4$ order \cite{JianCM2024,ZhangYH2024a}), the free bosons, $\beta$, and $\gamma$ phases all yield $G=\frac{3}{2}\frac{e^2}{h}$, and these phases occupy a significant portion of the phase diagram as plotted in Fig.~\ref{Fig:PD_n2}. The edge state of the putative FTI belongs to one of these three phases. One way to distinguish the three phases is to apply an in-plane magnetic field and monitor the response (see Sec.~\ref{Sec:B_field} for a detailed discussion). In particular, the free bosons phase with a finite magnetic field becomes a partially localized phase with a quantized edge-state conductance $G=\frac{1}{2}\frac{e^2}{h}$, which is a concrete, verifiable prediction based on our theory.
	
	Besides the putative FTI state, the twisted MoTe$_2$ experiment \cite{KangK2024c} showed an edge-state conductance close to $2\frac{e^2}{h}$ at $\nu_{\text{tot}}=4$, which was interpreted as the existence of double quantum spin Hall effect. A twisted WSe$_2$ experiment \cite{KangK2024a} from the same group also reported the signatures of double quantum spin Hall effect. The main idea is that the system has a nearly perfect $S_z$ conservation, suppressing any spin-flipping processes. The generic TR symmetric Anderson localization is parametrically weakened (i.e.,  the localization length is larger than or comparable to the edge length), resulting in a nearly quantized edge-state transport. 
	However, it is unnecessary to assume a nearly perfect $S_z$ conservation for the nearly ballistic conduction of two hLLs. In Fig.~\ref{Fig:PD_n1}, we show that $G=2\frac{e^2}{h}$ can happen in the free bosons, $\beta$, and $\gamma$ phases without assuming exactly vanishing Rashba spin-orbit coupling. Further experiments are needed to identify the precise phase of the two hLLs.

	We conclude this work by pointing out several future directions. In this work, we focus on the simplest Abelian composite helical edge states that can be described by two helical liquids. It is interesting to study all the Abelian composite helical edge states, including those with three or more helical liquids. A systematic framework that generalizes the studies for FQH edges \cite{MooreJE1998,MooreJE2002} might be needed. To understand the putative FTI in twisted MoTe$_2$ \cite{KangK2024c}, it is important to investigate the edge states of other possible FTI and provide experimentally relevant results, such as edge-state conductance and magnetic field response. Thus, extending the analysis of this work to possible non-Abelian states (e.g., the product Pfaffian states \cite{May-MannJ2024} and the weak-pairing phase in Ref.~\cite{ZhangYH2024a}) is highly desirable. Finally, future theoretical and experimental studies should characterize the primary sources of scatterings in the helical edge state specific to the moir\'e transition metal dichalcogenides, e.g., establishing temperature dependence of conductance and investigating the role of twist-angle disorder to the edge state.

	\begin{acknowledgments}
		We are grateful to Yahui Zhang for many useful suggestions to an earlier version of the manuscript and for explaining his work \cite{ZhangYH2024,ZhangYH2024a}.
		We also thanks Jay D. Sau for the discussions on the two-terminal conductance. This work is supported by the Laboratory for Physical Sciences.
	\end{acknowledgments}
	
	\appendix
	
	\begin{widetext}

		\section{Mapping to symmetric and antisymmetric sectors}\label{App:Mapping_SA}

		In this section, we discuss how to simplify the bosonic theory given by Eq.~(\ref{Eq:S_n'}), providing a convenient basis for evaluating the scaling dimensions. The goal is to map Eq.~(\ref{Eq:S_n'}) to a standard coupled two hLLs problem with symmetric and antisymmetric sectors \cite{KlesseR2000,ChouYZ2015,ChouYZ2019}.

		The action in Eq.~(\ref{Eq:S_n'}) can be expressed by
		\begin{align}
			\mathcal{S}^{(n)}_{2\text{hLL}}=&\frac{1}{4\pi}\int d\tau dx\left\{\begin{array}{c}
				\left(\partial_{x}\tilde{\varphi}_{1R}\right)\left(i\partial_{\tau}\tilde{\varphi}_{1R}\right)-\left(\partial_{x}\tilde{\varphi}_{1L}\right)\left(i\partial_{\tau}\tilde{\varphi}_{1L}\right)+\left(\partial_{x}\tilde{\varphi}_{2R}\right)\left(i\partial_{\tau}\tilde{\varphi}_{2R}\right)-\left(\partial_{x}\tilde{\varphi}_{2L}\right)\left(i\partial_{\tau}\tilde{\varphi}_{2L}\right)\\[2mm]
				+v_1\left[\left(\partial_x\tilde{\varphi}_{1R}\right)^2+\left(\partial_x\tilde{\varphi}_{1L}\right)^2\right]-2v_1'\left(\partial_x\tilde{\varphi}_{1R}\right)\left(\partial_x\tilde{\varphi}_{1L}\right)\\[2mm]
				+v_2\left[\left(\partial_x\tilde{\varphi}_{2R}\right)^2+\left(\partial_x\tilde{\varphi}_{2L}\right)^2\right]
				-2v_2'\left(\partial_x\tilde{\varphi}_{2R}\right)\left(\partial_x\tilde{\varphi}_{2L}\right)\\[2mm]
				+2u\left[\left(\partial_x\tilde{\varphi}_{1R}\right)\left(\partial_x\tilde{\varphi}_{2R}\right)+\left(\partial_x\tilde{\varphi}_{1L}\right)\left(\partial_x\tilde{\varphi}_{2L}\right)\right]
				-2u'\left[\left(\partial_x\tilde{\varphi}_{1R}\right)\left(\partial_x\tilde{\varphi}_{2L}\right)+\left(\partial_x\tilde{\varphi}_{1L}\right)\left(\partial_x\tilde{\varphi}_{2R}\right)\right]
			\end{array}
			\right\}.
		\end{align}
		Using $\tilde{\varphi}_{aR}=\phi_a+ \theta_a$ and $\tilde{\varphi}_{aL}=\phi_a-\theta_a$, the expression becomes
		\begin{align}
			\mathcal{S}^{(n)}_{2\text{hLL}}\rightarrow&\frac{1}{4\pi}\int d\tau dx\left\{\begin{array}{c}
				4\left(\partial_{x}\theta_1\right)\left(i\partial_{\tau}\phi_1\right)+4\left(\partial_{x}\theta_2\right)\left(i\partial_{\tau}\phi_2\right)\\[2mm]
				+2v_1\left[\left(\partial_x\phi_1\right)^2+\left(\partial_x\theta_1\right)^2\right]-2v_1'\left[\left(\partial_x\phi_1\right)^2-\left(\partial_x\theta_1\right)^2\right]\\[2mm]
				+2v_2\left[\left(\partial_x\phi_2\right)^2+\left(\partial_x\theta_2\right)^2\right]-2v_2'\left[\left(\partial_x\phi_2\right)^2-\left(\partial_x\theta_2\right)^2\right]\\[2mm]
				+4u\left[\left(\partial_x\phi_1\right)\left(\partial_x\phi_2\right)+\left(\partial_x\theta_1\right)\left(\partial_x\theta_2\right)\right]-4u'\left[\left(\partial_x\phi_1\right)\left(\partial_x\phi_2\right)-\left(\partial_x\theta_1\right)\left(\partial_x\theta_2\right)\right]
			\end{array}
			\right\}\\
			=&\int d\tau dx\left\{\begin{array}{c}
				\frac{1}{\pi}\left(\partial_{x}\theta_1\right)\left(i\partial_{\tau}\phi_1\right)+\frac{1}{\pi}\left(\partial_{x}\theta_2\right)\left(i\partial_{\tau}\phi_2\right)\\[2mm]
				+\frac{1}{2\pi}\left[\left(v_1-v_1'\right)\left(\partial_x\phi_1\right)^2+\left(v_1+v_1'\right)\left(\partial_x\theta_1\right)^2\right]\\[2mm]
				+\frac{1}{2\pi}\left[\left(v_2-v_2'\right)\left(\partial_x\phi_2\right)^2+\left(v_2+v_2'\right)\left(\partial_x\theta_2\right)^2\right]\\[2mm]
				+\frac{1}{\pi}\left[(u-u')\left(\partial_x\phi_1\right)\left(\partial_x\phi_2\right)+(u+u')\left(\partial_x\theta_1\right)\left(\partial_x\theta_2\right)\right]
			\end{array}
			\right\}.
		\end{align}
		Then, we rewrite the above expression with the collective variables, $\phi_{\pm}$ and $\theta_{\pm}$.
		\begin{align}
			\mathcal{S}^{(n)}_{2\text{hLL}}=&\int d\tau dx\left\{\begin{array}{c}
				\frac{1}{\pi}\left(\partial_{x}\theta_+\right)\left(i\partial_{\tau}\phi_+\right)+\frac{1}{\pi}\left(\partial_{x}\theta_-\right)\left(i\partial_{\tau}\phi_-\right)\\[2mm]
				+\frac{1}{2\pi}\frac{(v_1+v_2-v_1'-v_2')}{2}\left[\left(\partial_x\phi_+\right)^2+\left(\partial_x\phi_-\right)^2\right]+\frac{1}{2\pi}\frac{(v_1+v_2+v_1'+v_2')}{2}\left[\left(\partial_x\theta_+\right)^2+\left(\partial_x\theta_-\right)^2\right]\\[2mm]
				+\frac{1}{2\pi}\left[(v_1-v_2-v_1'+v_2')\left(\partial_x\phi_+\right)\left(\partial_x\phi_-\right)+(v_1-v_2+v_1'-v_2')\left(\partial_x\theta_+\right)\left(\partial_x\theta_-\right)\right]\\[2mm]
				+\frac{1}{2\pi}(u-u')\left[\left(\partial_x\phi_+\right)^2-\left(\partial_x\phi_-\right)^2\right]
				+\frac{1}{2\pi}(u+u')\left[\left(\partial_x\theta_+\right)^2-\left(\partial_x\theta_-\right)^2\right]
			\end{array}
			\right\}\\
			\label{Eq:S_n_+-}=&\int d\tau dx\left\{\begin{array}{c}
				\frac{1}{\pi}\left(\partial_{x}\theta_+\right)\left(i\partial_{\tau}\phi_+\right)+\frac{v_+}{2\pi}\left[K_+\left(\partial_x\phi_+\right)^2+\frac{1}{K_+}\left(\partial_x\theta_+\right)^2\right]\\[2mm]
				+\frac{1}{\pi}\left(\partial_{x}\theta_-\right)\left(i\partial_{\tau}\phi_-\right)+\frac{v_-}{2\pi}\left[K_-\left(\partial_x\phi_-\right)^2+\frac{1}{K_-}\left(\partial_x\theta_-\right)^2\right]\\[2mm]
				+\frac{1}{2\pi}\left[w\left(\partial_x\phi_+\right)\left(\partial_x\phi_-\right)+w'\left(\partial_x\theta_+\right)\left(\partial_x\theta_-\right)\right]
			\end{array}
			\right\},
		\end{align}
		where $v_+$ and $K_+$ ($v_-$ and $K_-$) correspond to the velocity and Luttinger parameter for the symmetric (antisymmetric) sector, $w=v_1-v_2-v_1'+v_2'$, and $w'=v_1-v_2+v_1'-v_2'$. $v_{\pm}$ and $K_{\pm}$ are related to the microscopic quantities through
		\begin{subequations}
			\begin{align}
				v_{\pm}K_{\pm}=&\frac{v_1+v_2-v_1'-v_2'}{2}\pm(u-u'),\\
				v_{\pm}/K_{\pm}=&\frac{v_1+v_2+v_1'+v_2'}{2}\pm(u+u').
			\end{align}
		\end{subequations}	
		Solving the above equations, we obtain
		\begin{align}
			v_{\pm}=&\left\{\left[\frac{v_1+v_2-v_1'-v_2'}{2}\pm(u-u')\right]\left[\frac{v_1+v_2+v_1'+v_2'}{2}\pm(u+u')\right]\right\}^{1/2},\\
			K_{\pm}=&\left\{\left[\frac{v_1+v_2-v_1'-v_2'}{2}\pm(u-u')\right]/\left[\frac{v_1+v_2+v_1'+v_2'}{2}\pm(u+u')\right]\right\}^{1/2},
		\end{align}
		
		In the main text, we consider a special limit that $v_1=v_2=v$ and $v_1'=v_2'=v'$. The $w$ and $w'$ in Eq.~(\ref{Eq:S_n_+-}) vanish in this limit. The action becomes two decoupled Luttinger liquids, described by Eq.~(\ref{Eq:S_+-_decoupled}). In this case, both $v_{\pm}$ and $K_{\pm}$ should be positive. Thus, the microscopic parameters need to satisfy
		\begin{align}
			(v-v')\pm(u-u')>0,\,\,\,\,(v+v')\pm(u+u')>0.
		\end{align}
		Without loss of generality, we choose $v>0$.
		The velocities and Luttinger parameters become
		\begin{align}
			v_{\pm}=&\sqrt{\left[(v-v)'\pm(u-u')\right]\left[(v+v')\pm(u+u')\right]},\\
			K_{\pm}=&\sqrt{\frac{(v-v')\pm(u-u')}{(v+v')\pm(u+u')}}.
		\end{align}
		For $v'=u'=0$, $v_{\pm}=v\pm u$ and $K_{\pm}=1$, suggesting that $u$ mainly controls the difference between $v_+$ and $v_-$, not the Luttinger parameters. Note that the value of $K_{\pm}$ is determined by both the intrachannel and the interchannel interactions. For example, $K_{-}>1$ can be realized by pure repulsive interactions, provided that the interchannel repulsion is stronger than the intrachannel repulsion. For example, $u'>v'\ge0$, $v>u'-v'$, and $u=0$ yield $K_-$ without requiring any microscopic interaction to be negative (i.e., attractive interaction).

	\end{widetext}

	\section{Bosonized operators}\label{App:Bosonization}
	
	Here, we provide the bosonized expressions of perturbations discussed in the main text. We also identify the integer-valued $m$ vectors that are used in the stability and compatibility analysis.
	
	First, we discuss the inter-channel TR symmetric backscattering. There are two distinct operators in $\mathcal{O}_{M,a}$, $L^{\dagger}_2R_1$ and $R^{\dagger}_2L_1$. We discuss these two operators separately. Using bosonization, we obtain
	\begin{align}
		L^{\dagger}_2R_1\sim&e^{im_M^T\Phi}=e^{i(\tilde\varphi_{1R}-\sqrt{n}\tilde\varphi_{2L})}\\
		=&e^{i\left[\frac{1+\sqrt{n}}{\sqrt{2}}\left(\phi_-+\theta_+\right)+\frac{1-\sqrt{n}}{\sqrt{2}}\left(\phi_++\theta_-\right)\right]},\\[2mm]
		R^{\dagger}_2L_1\sim&e^{im_M'^T\Phi}=e^{i(\tilde\varphi_{1L}-\sqrt{n}\tilde\varphi_{2R})}\\
		=&e^{i\left[\frac{1+\sqrt{n}}{\sqrt{2}}\left(\phi_--\theta_+\right)+\frac{1-\sqrt{n}}{\sqrt{2}}\left(\phi_+-\theta_-\right)\right]},
	\end{align}
	where $m_M=[1, 0, 0, -n]^T$ and $m_M'=[0, 1, -n,0]^T$.

	The inter-channel backscattering interactions are described by the $\mathcal{O}_+$ and $\mathcal{O}_-$. The corresponding operators can be bosonized as follows:
	\begin{align}
		L_1^{\dagger}R_1L_2^{\dagger}R_2\sim&e^{im_+^T\Phi}=e^{i(\tilde\varphi_{1R}-\tilde\varphi_{1L}+\sqrt{n}\tilde\varphi_{2R}-\sqrt{n}\tilde\varphi_{2L})}\\
		=&e^{i\sqrt{2}\left[(1+\sqrt{n})\theta_++(1-\sqrt{n})\theta_-\right]},\\[2mm]
		L_1^{\dagger}R_1R_2^{\dagger}L_2\sim&e^{im_-^T\Phi}=e^{i(\tilde\varphi_{1R}-\tilde\varphi_{1L}-\sqrt{n}\tilde\varphi_{2R}+\sqrt{n}\tilde\varphi_{2L})}\\
		=&e^{i\sqrt{2}\left[(1+\sqrt{n})\theta_-+(1-\sqrt{n})\theta_+\right]},
	\end{align} 
	where $m_+=[1,-1,n,-n]^T$ and $m_-=[1,-1,-n,n]^T$.

	The intra-channel backscattering interactions can be expressed by 
	\begin{align}
		:(L^{\dagger}_1R_1)^2:\sim&e^{im_{2p,1}^T\Phi}=e^{i(2\tilde\varphi_{1R}-2\tilde\varphi_{1L})}\\
		=&e^{i2\sqrt{2}(\theta_++\theta_-)},\\[2mm]
		:(L^{\dagger}_2R_2)^2:\sim&e^{im_{2p,2}^T\Phi}=e^{i(2\sqrt{n}\tilde\varphi_{2R}-2\sqrt{n}\tilde\varphi_{2L})}\\
		=&e^{i2\sqrt{2n}(\theta_+-\theta_-)},
	\end{align}
	where $m_{2p,1}=[2,-2,0,0]^T$ and $m_{2p,2}=[0,0,2n,-2n]^T$.

	The inter-channel Josephson coupling can be bosonized as follows:
	\begin{align}
		R_2^{\dagger}L_2^{\dagger}L_1R_1\sim&e^{im_J^T\Phi}=e^{i(\tilde\varphi_{1R}+\tilde\varphi_{1L}-\sqrt{n}\tilde\varphi_{2R}-\sqrt{n}\tilde\varphi_{2L})}\\
		=&e^{i\sqrt{2}\left[\left(1+\sqrt{n}\right)\phi_-+\left(1-\sqrt{n}\right)\phi_+\right]},
	\end{align}
	where $m_J=[1,1,-n,-n]^T$.
	
	We also discuss the intrachannel single-electron backscattering induced by an in-plane magnetic field. The single-electron backscattering term is given by
	\begin{align}
		L^{\dagger}_1R_1\sim& e^{im_{B,1}^T\Phi}=e^{i(\tilde\varphi_{1R}-\tilde\varphi_{1L})}\\
		=&e^{i\sqrt{2}\left(\theta_++\theta_-\right)},\\
		L^{\dagger}_2R_2\sim& e^{im_{B,2}^T\Phi}=e^{i\sqrt{n}(\tilde\varphi_{2R}-\tilde\varphi_{2L})}\\
		=&e^{i\sqrt{2n}\left(\theta_+-\theta_-\right)},
	\end{align}
	where $m_{B,1}=[1,-1,0,0]^T$ and $m_{B,2}=[0,0,n,-n]^T$.
	
	\section{Checking stability and compatibility conditions}\label{App:Stability}
	
	Here, we discuss the stability of various perturbations by checking the null vector criteria \cite{HaldaneFDM1995}. The compatibility conditions of having multiple perturbations are also examined. First, we compute the lower bound of the scaling dimensions for the various operators, using the $m$ vectors obtained in the previous section. The results are summarized as follows:
	\begin{align}
		\mathcal{D}_M=&\left|\frac{1}{2}m_M^T\hat{K}^{-1}m_M\right|=\left|\frac{1}{2}m_M'^T\hat{K}^{-1}m_M'\right|=\frac{n-1}{2},\\
		\mathcal{D}_+=&\left|\frac{1}{2}m_+^T\hat{K}^{-1}m_+\right|=0,\\
		\mathcal{D}_-=&\left|\frac{1}{2}m_-^T\hat{K}^{-1}m_-\right|=0,\\
		\mathcal{D}_{2p,1}=&\left|\frac{1}{2}m_{2p,1}^T\hat{K}^{-1}m_{2p,1}\right|=0,\\	
		\mathcal{D}_{2p,2}=&\left|\frac{1}{2}m_{2p,2}^T\hat{K}^{-1}m_{2p,2}\right|=0,\\
		\mathcal{D}_J=&\left|\frac{1}{2}m_{J}^T\hat{K}^{-1}m_{J}\right|=0,\\
		\mathcal{D}_B=&\left|\frac{1}{2}m_{B,1}^T\hat{K}^{-1}m_{B,1}\right|=\left|\frac{1}{2}m_{B,2}^T\hat{K}^{-1}m_{B,2}\right|=0.
	\end{align}
	The above results indicate that all perturbations except $\mathcal{O}_{M,a}$ with $n>1$ can induce instability to the free boson action, removing a combination of modes indicated by the $m$ vector. The results of $\mathcal{D}_{\mathcal{X}}$ can also be obtained by minimizing the $\Delta_{\mathcal{X}}$ with respect to $K_+$ and $K_-$. The derivation here is universal and independent of the $\hat{V}$ (or equivalently $\hat{\tilde{V}}$) matrix.
	
	In addition to the null-vector criteria for stability, we also check the compatibility condition in the following. Two vertex operators (constructed by $m$ and $m'$) can coexist in the strong coupling limit if $\mathcal{C}(m,m')\equiv |m^T\hat{K}^{-1}m'|=0$. We summarize the results in the following. First, we find
	\begin{align}
		\mathcal{C}(m_M,m_M')=&0,\\
		\mathcal{C}(m_+,m_-)=&0,\\
		\mathcal{C}(m_\pm,m_{2p,a})=&0,\\
		\mathcal{C}(m_{2p,1},m_{2p,2})=&0,\\
		\mathcal{C}(m_{B,1},m_{B,2})=&0,\\
		\mathcal{C}(m_{B,a},m_{\pm})=&0,\\
		\mathcal{C}(m_{B,a},m_{2p,a'})=&0,
	\end{align}
	indicating the pairs of perturbations can coexist in the strong coupling limit. The nontrivial $\mathcal{C}$'s are as follows:
	\begin{align}
		\mathcal{C}(m_M,m_+)=&\mathcal{C}(m_M',m_+)=n-1,\\
		\mathcal{C}(m_M,m_-)=&\mathcal{C}(m_M',m_-)=n+1,\\
		\mathcal{C}(m_M,m_{2p,1})=&\mathcal{C}(m_M',m_{2p,1})=2,\\
		\mathcal{C}(m_M,m_{2p,2})=&\mathcal{C}(m_M',m_{2p,2})=2n,\\
		\mathcal{C}(m_M,m_{J})=&\mathcal{C}(m_M',m_{J})=n-1,\\
		\mathcal{C}(m_M,m_{B,1})=&\mathcal{C}(m_M',m_{B,1})=1,\\
		\mathcal{C}(m_M,m_{B,2})=&\mathcal{C}(m_M',m_{B,2})=n,\\
		\mathcal{C}(m_+,m_J)=&2n-2,\\
		\mathcal{C}(m_-,m_J)=&2n+2,\\
		\mathcal{C}(m_{2p,1},m_J)=&4,\\
		\mathcal{C}(m_{2p,2},m_J)=&4n,\\
		\mathcal{C}(m_{B,1},m_J)=&2,\\
		\mathcal{C}(m_{B,2},m_J)=&2n,
	\end{align}
	The above results imply competitions between the pairs of vertex operators (corresponding to the pairs of $m$ vectors) in the strong coupling limit. A special situation is $n=1$, in which $\mathcal{C}(m_M,m_+)$, $\mathcal{C}(m_M',m_+)$, $\mathcal{C}(m_M,m_J)$, $\mathcal{C}(m_M',m_J)$, and $\mathcal{C}(m_+,m_J)$ are reduced to zero.

	\section{Excitation}\label{App:excitation}
	
	We discuss the charge excitations in an insulating edge state. There are three types of insulators: A localized insulator due to an applied in-plane magnetic field, a trivial Anderson insulator governed by $\mathcal{O}_{M,1}$ and $\mathcal{O}_{M,2}$ for the $n=1$ case, and a TRSB localized insulator governed by at least two out of $\mathcal{O}_+$, $\mathcal{O}_-$, and $\mathcal{O}_{2p,1}$. 
	
	First, we consider the trivial filed-induced localization. The dominant backscattering perturbations are $L^{\dagger}_1R_1\sim e^{i(\varphi_{1R}-\varphi_{1L})}$ and $L^{\dagger}_2R_2\sim e^{in(\varphi_{1R}-\varphi_{1L})}$. We focus only on the second channel as the first channel is the same as the $n=1$ case in the second channel. A local excitation corresponds to
	\begin{align}
		\delta\left(n\varphi_{2R}-n\varphi_{2L}\right)=2\pi P,
	\end{align}
	where $P$ is an integer. We arrive that $\delta N_2=\frac{1}{2\pi}\delta(\varphi_{2R}-\varphi_{2L})=P/n$, indicating a minimal $e/n$ charge excitation. For $n=1$, the localized state is an Anderson insulator with a minimal charge $e$ excitation.

	Then, we discuss the trivial Anderson insulator in the $n=1$ case. The dominant perturbations are $\mathcal{O}_{M,1}$ and $\mathcal{O}_{M,2}$, which induce single-particle backscattering. In bosonizations, these two perturbations are related to vertex operators $e^{i(\varphi_{1R}-\varphi_{2L})}$ and $e^{i(\varphi_{1L}-\varphi_{2R})}$. A local excitation corresponds to 
	\begin{align}
		\delta(\varphi_{1R}-\varphi_{2L})=&2\pi P_1,\\
		\delta(\varphi_{1L}-\varphi_{2R})=&2\pi P_2,
	\end{align}
	where $P_1$ and $P_2$ are integers. Using TR symmetry and the above equations, we derive the change of charge for each channel $\delta N_1=\frac{1}{2\pi}\delta\left(\varphi_{1R}-\varphi_{1L}\right)=P_1$ and $\delta N_2=\frac{1}{2\pi}\delta\left(\varphi_{2R}-\varphi_{2L}\right)=P_1$.
	The minimal charge excitation is $e$, consistent with a noninteracting Anderson insulator ($K_+=K_-=1$ in our theory).
	
	Finally, we discuss the TRSB localized insulator, which can be realized when both $\mathcal{O}_+$ and $\mathcal{O}_-$ perturbations become infinitely strong. A local excitation corresponds to
	\begin{align}
		\delta\left(\varphi_{1R}-\varphi_{1L}+n\varphi_{2R}-n\varphi_{2L}\right)=&2\pi Q_1,\\
		\delta\left(\varphi_{1R}-\varphi_{1L}-n\varphi_{2R}+n\varphi_{2L}\right)=&2\pi Q_2,
	\end{align} 
	where $Q_1$ and $Q_2$ are integers. We can easily derive the change of charge for each channel $\delta N_1=\frac{1}{2\pi}\delta\left(\varphi_{1R}-\varphi_{1L}\right)=(Q_1+Q_2)/2$ and $\delta N_2=\frac{1}{2\pi}\delta\left(\varphi_{2R}-\varphi_{2L}\right)=(Q_1-Q_2)/(2n)$. Thus, the minimal charge excitation is $e/(2n)$, half of the minimal charge of the composite edge state. The same conclusion can be obtained by considering other combinations of perturbations (e.g., $\mathcal{O}_{2p,a}$ and $\mathcal{O}_{\pm}$).

	\section{Two-terminal conductance}\label{App:Conductance}

\begin{figure}[t!]
		\includegraphics[width=0.45\textwidth]{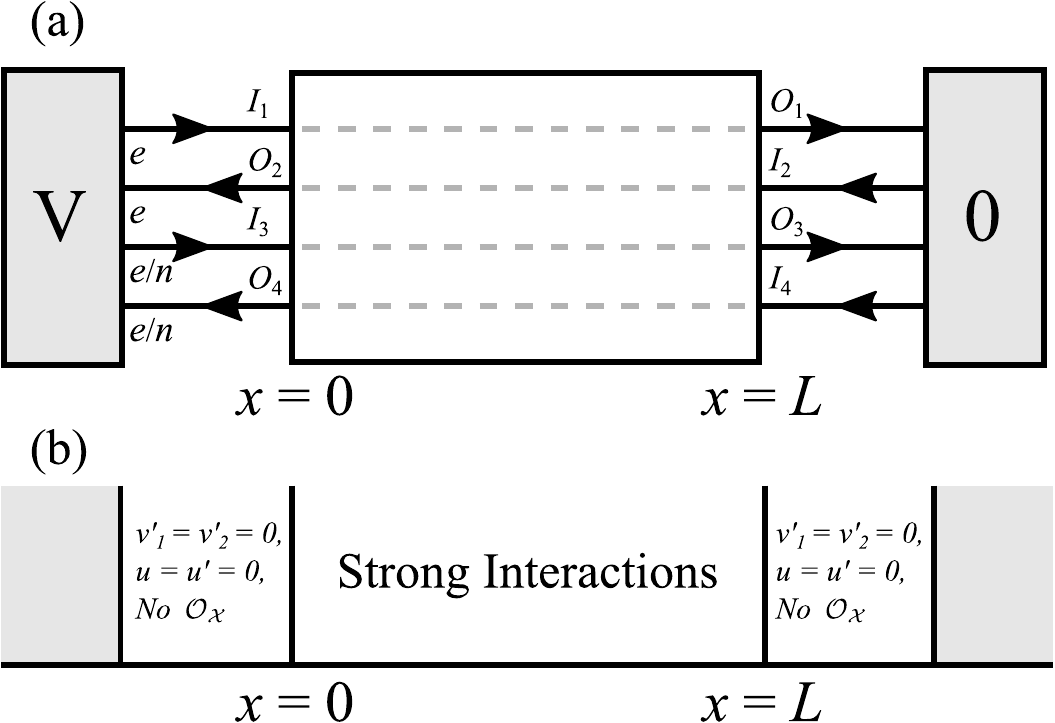}
		\caption{Effective setup of two-terminal edge-state conductance. (a) The ``strongly interacting'' regime (the central box between $x=0$ and $x=L$) is coupled to two reservoirs with voltages $0$ (right) and $V$ (left) through the ballistic channels (assuming screening of certain interactions). The top two chiral wires represent the regular hLL carrying charge $e$; the bottom two wires represent the helical liquid carrying charge $e/n$. These chiral wires indicate how the edge states couple to the external leads. The incoming ($I_j$) and outgoing ($O_j$) currents are also indicated. (b) The interaction parameters in the setup. For $x<0$ and $x>L$, the $\mathcal{O}_{\mathcal{X}}$ perturbations are absent, and the forward scattering interactions $v_1'=v_2'=u=u'=0$. For $0<x<L$, both the $\mathcal{O}_{\mathcal{X}}$ and the forward scattering interactions are generically present. Note that all the interactions are screened in the region of the reservoirs (gray regions).
		}
		\label{Fig:2term}
	\end{figure}

	We discuss the derivation of two-terminal conductance in this section, following Refs.~\cite{OregY2014,CornfeldE2015,ShavitG2019}. We model the system (one edge of the TI sample) by a central region with interaction and disorder connected to the reservoirs through ballistic chiral ``wires'' as shown in Fig.~\ref{Fig:2term}. First, we discuss how to realize the channels with fractionalized charges (wires 3 and 4). The fractionalized channels are the edge states of topological orders, which can be constructed systematically using the coupled-wire models \cite{KaneCL2002,NeupertT2014,TeoJCY2014,KlinovajaJ2014a}. The fractional conductance can be obtained explicitly using the Oreg-Sela-Stern formalism and taking the two-dimensional limit as discussed in Refs.~\cite{OregY2014,CornfeldE2015,ShavitG2019}. In the Oreg-Sela-Stern formalism, the interactions responsible for the charge fractionalization are ``screened out'' inside the leads, but the conductance is still fractionalized due to the collective partial gapping in the interacting region. Thus, we also assume that the backscattering interactions are completely absent in the lead regions.

	After establishing the fractionalized conduction channels, we consider a spatially varying $\hat{V}$ [given by Eq.~(\ref{Eq:V_matrix})] such that only $v_1$ and $v_2$ are left at $x=0$ and $x=L$. Physically, one can view this setup as the forward scattering interactions (i.e., $v_1'$, $v_2'$, $u$, and $u'$ in $\hat{V}$) and the $\mathcal{O}_{\mathcal{X}}$ are screened out by the reservoirs (leads) for $0<x<L$. For conceptual convenience, we further assume that the backscattering interactions responsible for charge fractionalization in wires 3 and 4 are much stronger than other perturbations, such that fractionalized channels (wires 3 and 4) survive in tiny regions outside the $0< x< L$ as shown in Fig.~\ref{Fig:2term}. The last assumption is unnecessary, but it helps us to simplify the setup and avoid possible confusion.

	Now, we discuss how to compute the conductance for the composite helical edge state. The ballistic wires are associated with the incoming ($O_j$) and outgoing ($O_j$) currents. First, we define a vector for the incoming currents, $J_{\text{in}}=[I_1,I_2,I_3,I_4]^T$, and a vector for the outgoing currents, $J_{\text{out}}=[O_1,O_2,O_3,O_4]^T$. The two vectors obey $J_{\text{out}}=\hat{S}J_{\text{in}}$, where $\hat{S}$ is a scattering matrix encoding perturbations. The incoming currents $I_1$ and $I_3$ ($I_2$ and $I_4$) come from the left (right) reservoir at potential $V$ (0). Thus, $J_{\text{in}}=\frac{e^2V}{h}[1,0,1/n,0]^T$. The $1/n$ factor here encodes the helical liquid with a charge $e/n$. The two-terminal conductance can be obtained via \cite{OregY2014}
	\begin{align}
		GV=&I_1+I_3-O_2-O_4=-I_2-I_4+O_1+O_3,\\
		\rightarrow G=&\frac{e^2}{h}\left[\begin{array}{cccc}
			1 & 0 & 1 & 0
		\end{array}\right]\hat{S}\left[\begin{array}{c}
			1\\
			0\\
			1/n\\
			0
		\end{array}\right].
	\end{align}
	In the absence of perturbation, the scattering matrix $\hat{S}=\text{diag}(1,1,1,1)$, yielding $G=\frac{e^2}{h}(1+1/n)$.
	Our goal is to construct the scattering matrix $\hat{S}$ for each case.
	
	If a perturbation satisfies the null-vector condition [i.e., $\mathcal{D}(m)=0$, where $m$ represents the corresponding vertex operator and $\mathcal{D}(m)$ is defined by Eq.~(\ref{Eq:D_m})], the combination of modes given by $m^T\Phi$ (the ``gapping'' mode) becomes a constant in time. As a result, $m^T\partial_t\Phi=0$, indicating a condition for the scattering matrix. The incoming and outgoing currents are related to the bosonic fields via
	\begin{align}
		I_1=&\frac{e}{2\pi}\partial_t\varphi_{1R}|_{x=0},\,\, O_1=\frac{e}{2\pi}\partial_t\varphi_{1R}|_{x=L},\\ 
		I_2=&\frac{e}{2\pi}\partial_t\varphi_{1L}|_{x=L},\,\, O_2=\frac{e}{2\pi}\partial_t\varphi_{1L}|_{x=0},\\ 
		I_3=&\frac{e}{2\pi}\partial_t\varphi_{2R}|_{x=0},\,\, O_3=\frac{e}{2\pi}\partial_t\varphi_{2R}|_{x=L},\\ 
		I_4=&\frac{e}{2\pi}\partial_t\varphi_{2L}|_{x=L},\,\, O_4=\frac{e}{2\pi}\partial_t\varphi_{2L}|_{x=0}.
	\end{align}
	Using the above equation and $m^T\partial_t\Phi|_{x=0,L}=0$, we can derive
	\begin{align}
		m_1I_1+m_2O_2+m_3I_3+m_4O_4=&0,\\
		m_1I_O+m_2I_2+m_3O_3+m_4I_4=&0.
	\end{align}
	In addition, there exist two ``ballistic'' modes, $w^T\Phi$ and $w'^T\Phi$, that are unaffected by the perturbation, satisfying $w^T\Phi|_{x=0}=w^T\Phi|_{x=L}$ and $w'^T\Phi|_{x=0}=w'^T\Phi|_{x=L}$. Thus, we derive two additional conditions
	\begin{align}
		w_1I_1\!+\!w_2O_2\!+\!w_3I_3\!+\!w_4O_4\!=&w_1O_1\!+\!w_2I_2\!+\!w_3O_3\!+\!w_4I_4,\\
		w'_1I_1\!+\!w'_2O_2\!+\!w'_3I_3\!+\!w'_4O_4\!=&w'_1O_1\!+\!w'_2I_2\!+\!w'_3O_3\!+\!w'_4I_4.
	\end{align}
	The choice of $w$ and $w'$ is not unique. The vectors $w$, $w'$, and $m$ need to be linearly independent. Generally, we can summarize the equations through $\hat{A}J_{\text{out}}=\hat{B}J_{\text{out}}$ and then derive $\hat{S}=\hat{A}^{-1}\hat{B}$.

	In the case of two compatible distinct perturbations satisfying the null-vector condition (i.e., everything is gapped), we can construct the $\hat{S}$ using conditions provided by the two gapping modes. However, this case is trivial, as the conductance is exactly zero when all the low-energy modes are removed.
	
	Now, we are in the position to derive the conductance of various cases in which some low-energy modes are removed due to perturbation. First, we discuss the $\mathcal{O}_+$ term, corresponding to $m_+=[1,-1,n,-n]^T$. The gapping conditions are given by
	\begin{align}
		I_1-O_2+nI_3-nO_4=&0,\\
		O_1-I_2+nO_3-nI_4=&0.
	\end{align}
	We choose $I_1=O_1$ and $I_3=O_3$ as the conditions for the ballistic modes. The scattering matrix is given by
	\begin{align}
		\hat{S}_+=\left[\begin{array}{cccc}
			1 & 0 & 0 & 0 \\
			0 & 1 & 0 & 0 \\
			-1/n & 1/n & 0 & 1 \\
			1/n & -1/n & 1 & 0 
		\end{array}\right].
	\end{align}
	The two-terminal conductance is $G=\frac{e^2}{h}(1-1/n)$. When $n=1$, the two-terminal conductance becomes zero, consistent with an antisymmetric fluid (after removing the symmetric sector) that does not respond to the external electric field. 
	
	We then discuss the $\mathcal{O_-}$ term, corresponding to $m_-=[1,-1,-n,n]^T$. The gapping conditions are given by
	\begin{align}
		I_1-O_2-nI_3+nO_4=&0,\\
		O_1-I_2-nO_3+nI_4=&0.
	\end{align}
	We choose $I_1=O_1$ and $I_3=O_3$ as the conditions for the ballistic modes. The scattering matrix is given by
	\begin{align}
		\hat{S}_-=\left[\begin{array}{cccc}
			1 & 0 & 0 & 0 \\
			0 & 1 & 0 & 0 \\
			1/n & -1/n & 0 & 1 \\
			-1/n & 1/n & 1 & 0 
		\end{array}\right].
	\end{align}
	The resulting two-terminal conductance is $G=\frac{e^2}{h}(1+1/n)$. 
	
	Finally, we discuss the $\mathcal{O}_J$ term, corresponding to $m_J=[1,1,-n,-n]^T$. The gapping conditions are given by
	\begin{align}
		I_1+O_1-nI_3-nO_4=&0,\\
		O_1+I_2-nO_3-nI_4=&0.
	\end{align}
	Again, we choose $I_1=O_1$ and $I_3=O_3$ as the conditions for the ballistic modes. The scattering matrix is given by
	\begin{align}
		\hat{S}_-=\left[\begin{array}{cccc}
			1 & 0 & 0 & 0 \\
			0 & 1 & 0 & 0 \\
			1/n & 1/n & 0 & -1 \\
			1/n & 1/n & -1 & 0 
		\end{array}\right].
	\end{align}
	The two-terminal conductance is $G=\frac{e^2}{h}(1+1/n)$.
	
	\section{Alternative derivation for two-terminal conductance}\label{App:Alt_cond}
	
	In this Appendix, we discuss an alternative way to compute the two-terminal conductance. The main idea is to inspect the localized/gapped mode due to a perturbation in the strong coupling, corresponding to a constraint in the physical currents of the two channels. In a fixed realization of disorder, we generally expect that $m^T\Phi(t,x)=\eta(x)$,
	where $m$ is the integer vector indicating the perturbation $\mathcal{O}\sim e^{im^T\Phi}$ and $\eta(x)$ is related to the corresponding disorder potential. As a result, a constraint $m^T\partial_t\Phi(t,x)=0$ can be derived. The current flow pattern can be obtained by analyzing such a constraint.
	
	The constraint of the $\mathcal{O}_+$ case is described by $\partial_t(\varphi_{1R}-\varphi_{1L}+n\varphi_{2R}-n\varphi_{2L})=0$. Equivalently, we obtain $j_1+nj_2=0$, where $j_a$ is the current of the $a$th channel. Thus, $j_2=-j_1/n$, suggests a negative drag situation. This negative drag condition naturally gives rise to $G=(1-1/n)e^2/h$, consistent with the result in Appendix~\ref{App:Conductance}. 
	
	Similarly, for $\mathcal{O_-}$, the constraint is given by $\partial_t(\varphi_{1R}-\varphi_{1L}-n\varphi_{2R}+n\varphi_{2L})=0$. We thus derive $j_2=j_1/n$, corresponding to a positive drag and a conductance $G=(1+1/n)e^2/h$. The $\mathcal{O}_J$ case gives the same conductance as the $\mathcal{O}_-$ because $\mathcal{O}_J$ and $\mathcal{O}_-$ have the same low-energy modes.

	%%\bibliography{FTI}	


\end{document}